\documentclass[12pt, draftclsnofoot, onecolumn]{IEEEtran}
\pdfoutput=1
\usepackage[noadjust]{cite}
\usepackage{amsmath}
\usepackage{array}
\usepackage{mdwmath}
\usepackage{amssymb}
\usepackage{algorithm}
\usepackage[noend]{algpseudocode}
\usepackage{eqparbox}
\usepackage{stfloats}
\usepackage{url}
\usepackage[pdftex]{graphicx}
\graphicspath{{../pdf/}{../jpeg/}}
\DeclareGraphicsExtensions{.pdf,.jpeg,.png}
\usepackage{epstopdf}
\usepackage{amsfonts}
\usepackage[tight,footnotesize]{subfigure}

\begin{document}

\title{\LARGE Asymptotic Max-Min SINR Analysis of Reconfigurable Intelligent Surface Assisted MISO Systems}
\author{Qurrat-Ul-Ain~Nadeem,~\IEEEmembership{Member,~IEEE,} Abla~Kammoun,~\IEEEmembership{Member,~IEEE,} Anas~Chaaban,~\IEEEmembership{Member,~IEEE,}
        M{\'e}rouane~Debbah,~\IEEEmembership{Fellow,~IEEE,} and ~ Mohamed-Slim~Alouini,~\IEEEmembership{Fellow,~IEEE}
\thanks{Q.-U.-A. Nadeem and A. Chaaban are with School of Engineering, The University of British Columbia, Kelowna, Canada.  (e-mail:  \{qurrat.nadeem, anas.chaaban\}@ubc.ca)}
\thanks{ A. Kammoun and M.-S. Alouini are with King Abdullah University of Science and Technology (KAUST), Thuwal, Makkah Province, Saudi Arabia. (e-mail: \{abla.kammoun,slim.alouini\}@kaust.edu.sa)}
\thanks{M. Debbah is  with  CentraleSup{\'e}lec, Gif-sur-Yvette, France and Mathematical and Algorithmic Sciences Lab, Huawei France R\&D, Paris, France (e-mail: merouane.debbah@huawei.com, merouane.debbah@centralesupelec.fr).}
}

\markboth{}%
{Shell \MakeLowercase{\textit{et al.}}: Bare Demo of IEEEtran.cls for Journals}
\maketitle

\vspace{-.7in}
\begin{abstract}
This work focuses on the downlink of a single-cell multi-user system in which a base station (BS) equipped with $M$ antennas communicates with $K$ single-antenna users through a reconfigurable intelligent surface (RIS) installed in the line-of-sight (LoS) of the BS. RIS is envisioned to offer unprecedented spectral efficiency gains by utilizing $N$ passive reflecting elements that induce phase shifts on the impinging electromagnetic waves to smartly reconfigure the signal propagation environment. We study the minimum signal-to-interference-plus-noise ratio (SINR) achieved by the optimal linear precoder (OLP), that maximizes the minimum SINR subject to a given power constraint for any given RIS phase matrix, for the cases where the LoS channel matrix between the BS and the RIS is of rank-one and of full-rank. In the former scenario,  the minimum SINR achieved by the RIS-assisted link is bounded by a quantity that goes to zero with $K$. For the high-rank scenario, we develop accurate deterministic approximations for the parameters of the asymptotically OLP, which are then utilized to optimize the RIS phase matrix. Simulation results show that RISs can outperform half-duplex relays with a small number of passive reflecting elements while large RISs are needed to outperform full-duplex relays. 
\end{abstract}

\vspace{-.2in}
\section{Introduction}
The spectral efficiency of wireless networks has greatly improved in the last decade thanks to various technological advances, including massive multiple-input multiple-output (MIMO), millimeter wave (mmWave) communication and ultra-dense deployments of small cells. However, all these technologies consume colossal amounts of energy and incur high hardware implementation costs \cite{issue}. For example: massive MIMO antenna arrays require one dedicated radio frequency (RF) chain per antenna element, which is prohibitive from cost and power consumption perspectives. Moreover, massive MIMO performance is known to suffer when the propagation environment exhibits poor scattering conditions, while communication at mmWave frequencies suffers from high path and penetration losses, resulting in signal blockages. To address the need for green and sustainable future cellular networks, researchers are looking into energy efficient techniques to improve the system performance and provide some control over the propagation environment. 

Among the very recent hardware technologies that promise significant reduction in the energy consumption of future wireless networks, while realizing unprecedented spectral efficiency gains belong the reconfigurable intelligent surfaces (RIS)s \cite{LIS, LIS1, LIS2, LIS_new1, mag_wu, LIS_CS,Renzo, MS_nature, RA, LIS5, RIS, RIS1, RIS2,mishra}. RIS is an array of nearly passive, low-cost, reflecting elements with reconfigurable parameters. With the help of a smart controller, each element independently introduces a phase shift on the impinging electromagnetic wave. By smartly adjusting the phase shifts induced by all elements, the RIS can achieve certain communication objectives, e.g., overcome unfavorable propagation conditions, increase the coverage area, while consuming very low energy due to the passive nature of the elements. 

Traditional reflecting surfaces have had variety of applications in radar and satellite communications but their use in terrestrial wireless communication systems was not considered earlier as they could not adapt the induced phases  with the time-varying channels that constitute the wireless propagation environments. However, recent advances in Micro-Electrical-Mechanical Systems (MEMS) and metamaterials have made real-time reconfiguration of reflecting surfaces possible using miniaturized circuits that can even be powered by energy harvesting modules, rendering the surface to be energy neutral \cite{harvest, Renzo}. The current implementations of RIS include reflect-arrays \cite{RA} and metasurfaces comprising of software defined metamaterials \cite{Renzo}. 

It is important to note that RIS differs significantly from other existing related technologies such as active wireless relay \cite{relay_p1} and active large intelligent surface (LIS) based massive MIMO \cite{LISA}. The former assists the BS-to-users communication by actively regenerating and retransmitting signals, thereby incurring additional power consumption. On the other hand, RIS does not use any active transmit modules like power amplifiers, but only relies on low-cost passive elements to reflect the received signal \cite{mag_wu, LIS_new1}. It does not apply any sophisticated signal processing algorithms and can operate in full-duplex mode without self-interference. LIS is a newly proposed wireless communication system, that can be viewed as an extension of massive MIMO but it  scales  up  beyond  the traditional  antenna  array  concept by envisioning the entire surface to be electromagnetically active and capable of transmitting.  The preliminary works envision LISs to offer  unprecedented data-transmission rates and show that as  long  as  the  distance  between  two  users  is larger  than  half  the  wavelength,  the  inter-user  interference  is negligible   \cite{LISA, LISAA, LISAAA, LIS_cc}. Most of these works study the information-transfer capabilities of the LIS through near-field analysis under perfect line-of-sight (LoS) propagation.  RIS has a completely different operating mechanism as it uses a passive surface for reflection and not transmission. 




The design of RIS parameters to realize various communication objectives is now attracting significant research interest, with more focus on indoor environments. In \cite{MS_nature}, the authors proposed to use an electronically tunable metasurface as a spatial microwave modulator. The metasurface deployed in a typical office room was shown to passively increase the wireless transmission between two antennas by an order of magnitude. The authors in \cite{RA} designed a reconfigurable reflect-array and studied its role in  establishing robust mmWave connections in indoor scenarios where the direct links are blocked. In \cite{LIS5} the authors envisioned the use of HyperSurfaces, which are software controlled metamaterials embedded in the environment, to interact with the electromagnetic waves in a fully software-defined fashion. 


A few recent works optimize the RIS parameters for outdoor communication scenarios \cite{LIS, LIS1, LIS2, LIS_new1, LIS_CS,RIS, RIS1, RIS2,mishra}. The authors in \cite{LIS} studied an RIS-assisted single-user multiple-input single-output (MISO) system and optimized the values of the induced phases so as to maximize the total received signal power at the user.  The works in \cite{LIS1, LIS2} proposed designs for both the transmit power allocation at the BS and the phases induced by the RIS elements to maximize either the energy or the spectral efficiency of an RIS-assisted multi-user MISO system, that employs zero-forcing precoding and models the RIS-assisted links as Rayleigh fading channels while the direct link is considered to be blocked. The authors in \cite{RIS} studied an RIS-assisted MIMO system by utilizing stochastic geometry tools to derived closed-form expressions for outage probability and users' ergodic rates. All these proposed designs require the RIS to have global CSI with respect to all the users as well as the BS, making the signal exchange overhead prohibitively high. 

In contrast to the previous literature, this paper is novel in three ways. First, it studies an RIS-assisted multi-user MISO system, with a LoS channel between the BS and the RIS and correlated Rayleigh  channels between the RIS and the users, taking into account the effects of both the rank and correlation structure of the channel. The design of MIMO precoding in literature is mostly based on two common optimization criteria - the transmit power minimization and the maximization of the minimum SINR \cite{MMS}. The latter has not been studied in the context of RIS parameters design and is the focus and second novelty of this work. Third, the proposed reflect beamforming design requires knowledge of only the channel large-scale statistics instead of the global CSI. The main contributions of this work can be summarized as follows.

\begin{itemize}
\item We propose a realistic signal model for the considered RIS-assisted MISO downlink communication system, where the RIS is installed in the direct LoS of the BS. The LoS assumption is practical since the RIS is usually deployed on the facade of a high rise building in the vicinity of the BS with knowledge of the BS’s location \cite{LIS}. As a result, the corresponding channel matrix between the BS and the RIS will likely be of rank one.
\item We formulate the problem of determining the optimal linear precoder (OLP), the power allocation matrix at the BS as well as the RIS phase shifts matrix that maximize the minimum SINR of the system. For any fixed RIS phase shifts matrix, the OLP and the optimal powers are obtained as the solution of a system of fixed-point equations \cite{MMS1}. 
\item For the rank-one LoS channel scenario, we show that the OLP reduces to maximum ratio transmission (MRT) and the minimum SINR can be expressed in a closed-form, bounded by a quantity that goes zero with the number of users. Therefore, it only makes sense to study the single-user setting for this scenario, for which we optimize the RIS phases. 
\item To benefit from the RIS in the multi-user setting, the BS-to-RIS LoS channel needs to have a high-rank. Assuming a full-rank BS-to-RIS channel matrix, we develop the deterministic approximations of the parameters of the OLP using random matrix theory (RMT) tools. 
\item Using the developed deterministic approximations, an efficient algorithm based on projected gradient ascent is proposed to solve the non-convex optimization problem of determining the phases that maximize the minimum user SINR under OLP. The developed algorithm requires knowledge of only the channel's large-scale statistics.
\item Simulation results compare the performance of the RIS with that of both half-duplex and full-duplex amplify-and-forward relays and provide useful insights into the impact of the number of reflecting elements and the channel rank on the performance.
\end{itemize} 

The rest of the paper is organized as follows. In Section II, the RIS-assisted MISO system is described, the max-min SINR problem is formulated and the OLP is reviewed. In Section III, the rank-one BS-to-RIS LoS channel scenario is analyzed. Section IV derives asymptotically tight deterministic equivalents of the OLP for the full-rank channel scenario and designs the RIS phases. Simulation results are provided in Section V and Section VI concludes the paper.

\section{System Model}

In this section, we present the transmission model for the considered RIS-assisted downlink multi-user MISO system. We also describe the targeted max min SINR problem formulation and recall the solution for the OLP and the optimal power allocation from \cite{MMS1}.

\begin{figure}
\centering
\includegraphics[width=3.25in, height=1.5in]{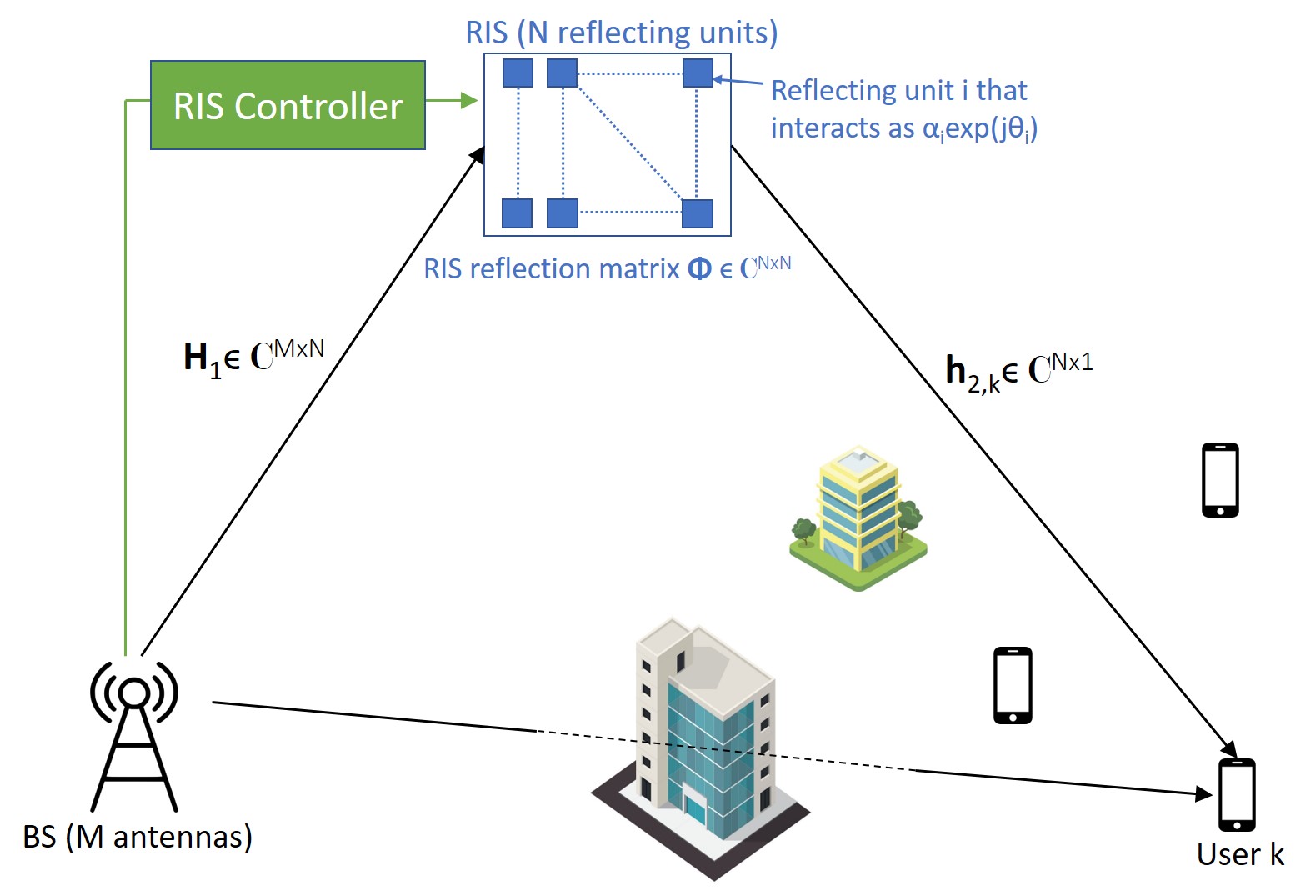}
\caption{RIS-assisted MISO system.}
\label{LIS_model}
\end{figure}

\vspace{-.1in}
\subsection{Signal Model}

As shown in Fig. \ref{LIS_model}, we consider a multi-user MISO communication system where the BS equipped with $M$ antennas serves $K$ single-antenna  users. An RIS composed of $N$ passive reflecting elements is installed on the wall of a surrounding high-rise building to assist the BS in communicating with the users. The RIS can dynamically adjust the phase shift induced by each reflecting element. Besides, we assume that the direct links from the BS to the users are blocked by obstacles, such as buildings. While such an assumption is more applicable in indoor communication scenarios, there are several works that study outdoor RIS-assisted communication systems under a blocked direct link \cite{LIS1, LIS2, LIS_new1, RIS, RIS1, RIS2}. This assumption is likely to be true for 5G and beyond mmWave and sub-mmWave communication systems, which are known to suffer from high path and penetration losses resulting in signal blockages \cite{mmwave11, relay_p1}.  Additionally, we consider a time-division duplex protocol and assume transmission over quasi-static flat-fading channels. The received baseband signal $y_{k}$ at user $k$ is given as,
\begin{align}
&y_{k}=\left(\sqrt{\beta_{k}}\textbf{h}_{2,k}^{H}\textbf{R}_{RIS_{k}}^{1/2}\boldsymbol{\Phi}^{H}\textbf{H}_{1}^{H}\right)\textbf{x}+n_{k},
\end{align}
where $\beta_{k}$ represent the channel attenuation coefficient of the RIS-assisted link for user $k$, $\textbf{H}_{1}\in\mathbb{C}^{M\times N}$ represents the LoS channel matrix between the BS and RIS, $\textbf{R}_{RIS_{k}} \in \mathbb{C}^{N\times N}$ represents the spatial correlation matrix for the RIS elements with respect to user $k$,  $\textbf{h}_{2,k} \sim \mathcal{CN}(\textbf{0},\textbf{I}_N)\in \mathbb{C}^{N\times 1}$ denotes the channel vector between the RIS and user $k$ and the term $n_{k}$ represents the thermal noise, modeled as a $\mathcal{CN}(0,\sigma^2)$ random variable (RV). In addition, $\boldsymbol{\Phi}=\text{diag}(\alpha \phi_{1}, \dots, \alpha \phi_{N}) \in \mathbb{C}^{N\times N}$ is a diagonal matrix accounting for the response of the RIS elements, where $\phi_{n}=\exp(j\theta_{n})$, $n=1,\dots, N$. Note that $\theta_{n} \in [0,2\pi]$ are the phase shifts applied by the RIS elements and $\alpha \in (0,1]$ is the fixed amplitude reflection coefficient. 

To ensure maximum signal reflection, existing works on RIS-assisted systems set $\alpha=1$, justifying this choice with the significant research progress made on the development of lossless metasurfaces that can achieve unity reflection magnitude with arbitrary reflection phase \cite{MS_loss, MS_loss2, MS_loss3, MS_loss1}. In practice, these metasurfaces  will  absorb some incident waves and produce  parasitic  reflections into unwanted directions, resulting in reduced reflection efficiency of the RIS. Despite of these losses, designs achieving $\alpha>0.9$ have been reported in the literature \cite{MS_loss2, MS_loss3}. We assume $\alpha$ to be a given fixed value and will study its impact on the performance later through simulations.

The Tx signal \textbf{x} is given as $\textbf{x}=\sum_{k=1}^{K} (\sqrt{p_{k}/K}) \textbf{g}_{k} s_{k}$, where $\textbf{g}_{k} \in \mathbb{C}^{M\times 1}$, $p_{k}$ and $s_{k} \sim \mathcal{CN}(0,1)$ are the precoding vector, signal power  and data symbol for user $k$ respectively. The Tx vector satisfies the average Tx power per user constraint as,
\begin{align}
\label{p_cons}
& \mathbb{E}[||\textbf{x}||^{2}]=\text{tr }(\textbf{P} \textbf{G}^{H} \textbf{G})/K \leq  P_{max},
\end{align}
where $P_{max}>0$ is the Tx power constraint at the BS, $\textbf{P}=\text{diag}(p_{1}, \dots, p_{K})$ and $\textbf{G}=[\textbf{g}_{1}, \textbf{g}_{2}, \dots,\textbf{g}_{K}] \in \mathbb{C}^{M\times K}$. The downlink SINR at user $k$ is defined as,
\begin{align}
\label{SINR}
& \gamma_{k}=\frac{\frac{p_{k}}{K} |\textbf{h}_{k}^{H} \textbf{g}_{k}|^{2}}{\sum_{i\neq k} \frac{p_{i}}{K} |\textbf{h}_{k}^{H} \textbf{g}_{i}|^{2}+ 1}, \\
\label{ch_ov}
& \text{where, }\textbf{h}_{k}=\sqrt{\rho_{k}} \textbf{H}_{1}\boldsymbol{\Phi}\textbf{R}_{RIS_{k}}^{1/2}\textbf{h}_{2,k}.
\end{align} 
 We have normalized the noise variance to unity by introducing $\rho_{k}=\beta_k/\sigma^2$. Note that the overall channel $\textbf{h}_k\sim \mathcal{CN}(\textbf{0}, \textbf{R}_{k})$, where $\textbf{R}_{k}=\rho_{k} \textbf{H}_{1}\boldsymbol{\Phi}\textbf{R}_{RIS_{k}} \boldsymbol{\Phi}^H \textbf{H}_{1}^H$. 

In order to optimize \textbf{G} and $\boldsymbol{\Phi}$, CSI is needed at both the BS and the RIS, which is assumed to be perfectly available in most prior  works. Obtaining this CSI, however, is a critical challenge due to  the hardware constraints on RIS elements of being nearly passive. The latter prohibits the use of receive RF chains at the RIS to enable sensing capability for channel estimation. While keeping the RIS passive, there are two main approaches to obtain CSI.   The  first   is  to  switch the RIS elements ON one-by-one and estimate  the RIS-assisted channels at the BS based on received pilot symbols from the users \cite{mishra}.  This yields large channel  training  overhead  when $N$ is large.  Second approach is to select the RIS reflection matrix from quantized codebooks via   beam  training. For example, the RIS can quickly sweep a pre-designed codebook to select the best beam based on the BS/user received training signal power. The  size  of  the   codebook  is  in  the order $N$, which leads to large complexity and real-time overhead. Methods to reduce this overhead through compressive sensing and deep learning techniques have appeared in \cite{LIS_CS}. Another practical challenge in the successful commercialization of this technology, is the potential increase in the latency of wireless networks under dynamic channel conditions, where updating the RIS reflection matrix at the pace of fast fading channels and sharing this information between the BS and RIS to synchronize their operation can result in significant real-time overhead and delays.

 In this work, we show that when $M$,$N$ and $K$ are large, the optimized $\boldsymbol{\Phi}$ can be computed using knowledge of only the spatial correlation matrices of the users, which depend on the channels' large scale statistics that vary very slowly and do not need to be estimated at the pace of fast-fading. The BS shares the optimized $\boldsymbol{\Phi}$ with the RIS controller after several coherence intervals,  resulting in a reduction in the signal exchange overhead and the time delays caused by BS-RIS synchronization. However, we assume perfect CSI to be available at the BS which computes the precoder. In practice, having perfect CSI is challenging especially in RIS-assisted systems, where a much higher number of links is involved. Our results serve as relevant performance upper bounds for realistic scenarios where CSI errors will be present.


\vspace{-.13in}
\subsection{Optimal Linear Precoder}

For any given $\boldsymbol{\Theta}$, the optimal precoding vectors and the optimal allocated powers, $\textbf{p}=[p_{1},\dots, p_{K}]^T$, are chosen as the solution of the following max-min SINR problem. 
\begin{subequations}
 \begin{alignat}{2} \textit{(P1)} \hspace{.35in}
&\!\max_{\textbf{P}, \textbf{G}} \hspace{.1in} \!\min_{k}        &\qquad& \gamma_{k} \label{P1}\\
&\text{subject to} &      & \frac{1}{K}\textbf{1}_{K}^{T}\textbf{p} \leq P_{max}, \hspace{.1in} ||\textbf{g}_{k}||=1, \forall k.
\end{alignat}
\end{subequations}

In \cite{MMS1}, it was shown that the OLP, $\textbf{G}^*$, that solves (\textit{P1}) takes the form,
\begin{align}
\label{G_opt}
\textbf{g}_{k}^*=\frac{\left(\sum_{i\neq k} \frac{q_{i}^*}{K}\textbf{h}_{i}\textbf{h}_{i}^{H}+\textbf{I}_{M}\right)^{-1}\textbf{h}_{k}}{||\left(\sum_{i\neq k} \frac{q_{i}^*}{K}\textbf{h}_{i}\textbf{h}_{i}^{H}+\textbf{I}_{M}\right)^{-1}\textbf{h}_{k}||},
\end{align}
where $q_{k}^*$s are obtained as the unique positive solution of the following fixed-point equations:  
\begin{align}
\label{q}
q_{k}^*=\frac{\tau^*}{\frac{1}{K}\textbf{h}_{k}^H\left(\sum_{i\neq k} \frac{q_{i}^*}{K}\textbf{h}_{i}\textbf{h}_{i}^{H}+\textbf{I}_{M}\right)^{-1}\textbf{h}_{k}},
\end{align}
with $\tau^*$ being the minimum SINR under OLP given by,
\begin{align}
\label{tau}
\tau^*=\frac{KP_{max}}{\sum_{k=1}^{K}\left(\frac{1}{K}\textbf{h}_{k}^H\left(\sum_{i\neq k} \frac{q_{i}^*}{K}\textbf{h}_{i}\textbf{h}_{i}^{H}+\textbf{I}_{M}\right)^{-1}\textbf{h}_{k}\right)^{-1}}.
\end{align}
The optimal power coefficients $p_{k}^*$s are such that $\gamma_{1}^*=\gamma_{2}^*=\dots = \gamma_{K}^*=\tau^*$\cite{MMS1}, with $\gamma_{k}^*=\frac{\frac{p^*_{k}}{K} |\textbf{h}_{k}^{H} \textbf{g}^*_{k}|^{2}}{\sum_{i\neq k} \frac{p^*_{i}}{K} |\textbf{h}_{k}^{H} \textbf{g}^*_{i}|^{2}+ 1}$. It turns out that $\textbf{p}^*=[p_{1}^*,\dots, p_{K}^*]^{T}$ can be obtained as,
\begin{align}
\label{P_opt}
&\textbf{p}^*=\left(\textbf{I}_{K}-\tau^*\textbf{D}\textbf{F}  \right)^{-1}\tau^*\textbf{D}\textbf{1}_{K},
\end{align}
where $\textbf{D}=\text{diag}\left(\frac{1}{\frac{1}{K}|\textbf{h}_{1}^{H}\textbf{g}^*_{1}|^{2}}, \dots, \frac{1}{\frac{1}{K}|\textbf{h}_{K}^{H}\textbf{g}^*_{K}|^{2}} \right)$ and $[\textbf{F}]_{k,i}=\frac{1}{K}|\textbf{h}_{k}^{H}\textbf{g}^*_{i}|^{2}$, if $k\neq i$ and $0$ otherwise.

\vspace{-.1in}
\subsection{RIS Design Problem Formulation}

We aim to design the RIS reflect beamforming matrix $\boldsymbol{\Phi}$, that maximize the minimum SINR under OLP. To make the targeted problem more tractable, we assume that the RIS is equipped with infinite resolution phase shifters and formulate the reflect beamforming design problem as,
\begin{subequations}
 \begin{alignat}{2} \textit{(P2)} \hspace{.35in}
&\!\max_{\boldsymbol{\Phi}}      &\qquad& \tau^* \label{P2}\\
&\text{subject to} &      & |\phi_{n}|=1, \hspace{.08in} n=1,\dots, N.\label{constraint3}
\end{alignat}
\end{subequations}
The optimization problem in (\ref{P2}) is non-convex due to the objective function $\tau^*$ being non-convex in $\boldsymbol{\Phi}$ and also due to the unit-modulus constraint in (\ref{constraint3}). In the sequel, we will analyze the expression of $\tau^*$ and solve \textit{(P2)} for the cases where the LoS channel $\textbf{H}_{1}$ is represented by a rank-one matrix and where it is given by a full-rank matrix. 



\vspace{-.12in}
\section{RIS Design for Rank-One $\textbf{H}_{1}$}

The RIS is envisioned to be deployed on a high rise building in the vicinity of the BS, rendering the BS-to-RIS channel to be dominated by the LoS link \cite{LIS}. Since this is a point-to-point LoS communication link, so the corresponding channel matrix $\textbf{H}_{1}$ is of rank one, i.e. $\textbf{H}_{1}=\textbf{a}\textbf{b}^{H}$. Intuitively, this implies that the degrees of freedom offered by the RIS-assisted link is equal to one and deploying an RIS would not yield any substantial benefit, unless the system serves a single user at one time. In this section, we will prove this by first analyzing the expressions of $\tau^*$ in (\ref{tau}) and the OLP in (\ref{G_opt}).   Later, we solve problem \textit{(P2)} for a single-user system.


\subsection{Analysis of the OLP}

As shown in Section II-B, the OLP is parametrized by the scalars $q_{k}$s, where $q_{k}$s need to be evaluated by an iterative procedure due to the fixed-point equations in (\ref{q}) and (\ref{tau}). This is a computationally demanding task especially when $M$ and $K$ are large since the matrix inversion operation in (\ref{q}) and (\ref{tau}) must be recomputed at every iteration. However we find that when $\textbf{H}_{1}$ is of rank-one and the channel is composed of the RIS-assisted link only, the OLP is MRT \cite{MMS1} and the optimal minimum SINR, $\tau^*$, has a simple closed-form expression stated in \textbf{Theorem 1}.

\textbf{Theorem 1.}  Under the setting of a rank-one channel between the BS and the RIS, the optimal precoding strategy is MRT, where the columns vectors of the  OLP take the form,
\begin{align}
\label{MRTT}
&\textbf{g}^*_{k}=\frac{\textbf{h}_{k}}{|| \textbf{h}_{k}||},
\end{align}
and the minimum SINR under OLP in (\ref{tau}) is given by,
\begin{align}
\label{tau2}
&\tau^*=\frac{P_{max}}{Z+P_{max}(K-1)}, \text{ where, }Z=\frac{1}{M}\sum_{k=1}^{K}\frac{1}{\rho_k \textbf{h}_{2,k}^{H}\textbf{R}_{RIS_k}^{1/2}\boldsymbol{\Phi}^{H}\textbf{b}\textbf{b}^{H} \boldsymbol{\Phi} \textbf{R}_{RIS_k}^{1/2} \textbf{h}_{2,k}}.
\end{align}

\textit{Proof:} The proof is detailed in Appendix A.




The proof of \textbf{Theorem 1} yields the following Corollary.

\textbf{Corollary 1.} Under the setting of \textbf{Theorem 1}, the minimum SINR in (\ref{tau2}) is bounded as,
\begin{align}
\label{bound1}
&\tau^* \leq \frac{1}{K-1}.
\end{align}

\textit{Proof:} Solving (\ref{alpha}) for $\alpha_{k}\geq 0$ yields the corollary. 

This corollary implies that as $K$ increases, the optimal minimum SINR achieved by the RIS-assisted link alone is upper bounded by a quantity that goes to zero irrespective of the values of $M$ and $N$. This motivates us to study the convergence behavior of the RV $\tau^*$ as $M,K$ grow large (denoted as $M,K\rightarrow \infty$), which is not easy since the RV $Z$ in the expression of $\tau^*$ involves the sum of independent but non-identically distributed RVs. To facilitate the asymptotic analysis, we bound $\tau^*$ by $\tau^{*u}$ that involves the sum of i.i.d. RVs. To do this, we study the distribution of $Z$.

Let $X_{k}=\rho_k \textbf{h}_{2,k}^{H}\textbf{R}_{RIS_{k}}^{1/2}\boldsymbol{\Phi}^{H}\textbf{b}\textbf{b}^{H} \boldsymbol{\Phi} \textbf{R}_{RIS_{k}}^{1/2} \textbf{h}_{2,k}$. Observe that $X_{k} \sim \frac{\rho_k \sigma^2_{X_{k}}}{2 } \chi_{2}^{2}$, where $\chi_{k}^{2}$ denotes the chi-square distribution with $k$ degrees of freedom and $\sigma_{X_{k}}^2=\textbf{b}^{H} \boldsymbol{\Phi}  \textbf{R}_{RIS_{k}} \boldsymbol{\Phi} ^{H} \textbf{b}$.

Let $Z=\frac{1}{M}\sum_{k=1}^{K} Y_{k}$, where $Y_{k}=\frac{1}{X_{k}} \sim \frac{2 }{\rho_{k} \sigma_{X_{k}}^2} \text{Inv-}\chi^2_2$, where $\text{Inv-}\chi^2_k$ denotes the inverse chi-square distribution with $k$ degrees of freedom. Then we can bound $\tau^*$ in (\ref{tau2}) as follows,
\begin{align}
\label{eq1}
&\tau^*\leq \tau^{*u}=\frac{P_{max}}{P_{max}(K-1)+\frac{2}{\rho_{max} \sigma^{2}_{X,max}} \frac{1}{M}\sum_{k=1}^{K} \tilde{Y}_{k}},
\end{align}
where $\tilde{Y}_{k} \sim \text{Inv-}\chi^2_2$, $\rho_{max}=\text{max}(\rho_{1}, \dots, \rho_{K})$ and $\sigma^{2}_{X,max}=N\text{max}\{\lambda_{max}(\boldsymbol{\Phi}\textbf{R}_{RIS_{1}}\boldsymbol{\Phi}^{H}), \dots, \\ \lambda_{max}(\boldsymbol{\Phi}\textbf{R}_{RIS_{K}}\boldsymbol{\Phi}^{H}) \}$, where $\lambda_{max}(\textbf{A})$ denotes the maximum eigenvalue of \textbf{A}. We will now study $K\tau^{*u}$ as $M,K \rightarrow \infty$.  The analysis will require the following Lemma.


\textbf{Lemma 1.} Given $\tilde{Y}_{k}$, $k=1,\dots, K$, are distributed as $\text{Inv-}\chi^2_2$, then [\cite{book_s} Theorem 1.8.1],
\begin{align}
&\frac{1}{M}\sum_{k=1}^{K}\tilde{Y}_{k} - c \log K \xrightarrow[M,K\rightarrow \infty]{d} G,
\end{align}
where $\xrightarrow[M,K\rightarrow \infty]{d}$ denotes convergence in distribution, $c$ is a constant that depends on the distribution of $\tilde{Y}_{k}$s and $G \sim S_{1}(1,0,0)$, where $S_{\alpha}(\sigma,\beta,\mu)$ denotes the alpha-stable distribution.



To study the asymptotic behavior of $K\tau^{*u}$, we require the following two assumptions.

\textbf{Assumption 1.} The coefficients $\rho_{k}$ satisfy: $0 < \text{lim inf }\rho_{k} \leq \text{lim sup } \rho_{k} < \infty$, $\forall k$.

\textbf{Assumption 2.} The vector $\textbf{b}$ and matrices $\textbf{R}_{RIS_k}$ $\forall k$, satisfy: $ \text{lim inf } ||\textbf{b}|| >0$ and $\text{lim inf } \\ \lambda_{min}(\textbf{R}_{RIS_k})>0$ $\forall k$, where $\lambda_{min}(\textbf{A})$ denotes the minimum eigenvalue of \textbf{A}.


\textbf{Theorem 2.} Under the setting of \textbf{Theorem 1} and \textbf{Assumption 1} and \textbf{Assumption 2}, $K\tau^{*u}$, where $\tau^{*u}$ is defined in (\ref{eq1}), converges as,
\begin{align}
&K\tau^{*u} \xrightarrow[M,K\rightarrow \infty]{p} 1,
\end{align}
where $\xrightarrow[M,K\rightarrow \infty]{p}$ denotes convergence in probability. 

\textit{Proof:} The proof of \textbf{Theorem 2} is postponed to Appendix B.

\textbf{Theorem 2} shows that the minimum SINR achieved by the RIS-assisted link under rank-one $\textbf{H}_1$ will go to zero as $K$ increases, no matter how many antennas or RIS elements we deploy. This is because the first leg of the communication constitutes the pinhole channel, limiting the degrees of freedom offered by the channel and the number of users that can be efficiently served, to one. To serve multiple users simultaneously, we need to introduce enough rank in $\textbf{H}_{1}$, i.e. rank$(\textbf{H}_{1}) \geq K$. One way to introduce rank  is to deploy multiple RISs each with an independent LoS channel with the BS, such that the overall channel is given as,
\begin{align}
\label{ch_ov1}
&\textbf{h}_{k}=\sum_{l=1}^{L} \sqrt{\rho_{l,k}} \textbf{H}_{1,l}\boldsymbol{\Phi}_{l}\textbf{h}_{2,l,k},
\end{align}
where $L$ is the total number of RISs, $\textbf{H}_{1,l}$ is the LoS channel between the BS and RIS $l$ given as $\textbf{a}_l \textbf{b}_l^H$, $\textbf{h}_{2,l,k} \sim \mathcal{CN}(\textbf{0},\textbf{R}_{RIS_{l,k}})$ is the fast fading channel vector between the RIS $l$ and user $k$, $\textbf{R}_{RIS_{l,k}}$ represents the spatial correlation matrix at RIS $l$ and $\boldsymbol{\Phi}_{l}$ represents the reflect beamforming matrix for RIS $l$.  Intuitively under this model, the channel matrix $\textbf{H}=[\textbf{h}_1, \dots, \textbf{h}_K] \in \mathbb{C}^{M\times K}$ will have upto $L$ non-zero eigenvalues and hence $L$ degrees of freedom. Such a system can therefore support upto $L$ users for simultaneous transmission. A preliminary analysis of the model in (\ref{ch_ov1}) under the assumption that  $\textbf{a}_{l}^H \textbf{a}_{l'}=0$, if $l\neq l'$ yields \textbf{Corollary 2}.

\textbf{Corollary 2.} Under the channel model in (\ref{ch_ov1}), $\tau^*$ is upper bounded as,
\begin{align}
&\tau^* \leq \sqrt{\frac{L}{K}} \sqrt{P_{max}} \sqrt{\underset{l,k}{\text{max}} \frac{\rho_{l,k}\textbf{h}_{2,l,k}^H \boldsymbol{\Phi}_{l}^H  \textbf{b}_l \textbf{b}_l^H \boldsymbol{\Phi}_{l}\textbf{h}_{2,l,k} \underset{k}{\text{max}} \left(\frac{1}{K} \textbf{h}_{k}^H \textbf{h}_k  \right)}{\frac{1}{K} \textbf{b}_l^H  \boldsymbol{\Phi}_{l}  \textbf{H}_{2,l[k]} \bar{\textbf{P}}_{l[k]} \textbf{H}_{2,l[k]}^H  \boldsymbol{\Phi}_{l}^H  \textbf{b}_l}},
\end{align}
where $\textbf{H}_{2,l[k]}$ is the $N\times K-1$ matrix of vectors $\textbf{h}_{2,l,i}$, $i=1,\dots, K$ with column $k$ removed and $\bar{\textbf{P}}_{l[k]}=\text{diag}(\rho_{l,1}, \dots, \rho_{l,k-1}, \rho_{l,k+1}, \dots, \rho_{l,K})$.

\textit{Proof.} The proof is provided at the end of Appendix A.

A study of the order of terms reveals that $\textbf{h}_{2,l,k}^H \boldsymbol{\Phi}_{l}^H  \textbf{b}_l \textbf{b}_l^H  \boldsymbol{\Phi}_{l}\textbf{h}_{2,l,k}$  and $\frac{1}{K} \textbf{b}_l^H  \boldsymbol{\Phi}_{l} \textbf{H}_{2,l, [k]} \bar{\textbf{P}}_{l[k]}  \textbf{H}_{2,l, [k]}^H \boldsymbol{\Phi}_{l}^H  \textbf{b}_l$ are both of order $O(N^2)$. Moreover $\frac{1}{K} \textbf{h}_{k}^H \textbf{h}_k  $ is of order $O(M/K)$. Therefore the upper bound is of order $O(\sqrt{\frac{L}{K} }\sqrt{\frac{M}{K}})$ instead of $O(1/K)$ as was the case in \textbf{Corollary 1}, implying that as long as $L$ is comparable to $K$, the bound does not grow to $0$ with $K$. Therefore by using $L$ RISs, each with an orthogonal rank-one channel with the BS, the system can serve upto $L$ users.

Extending the  analysis in the subsequent sections to multiple RISs is out of the scope of this work as it renders the optimization problems much harder to solve and has been left as an important future direction to look into. Note that almost all existing works on RIS-assisted systems work with a single RIS as well. 


\subsection{Optimization of $\boldsymbol{\Phi}$ for a Single-User Setting}

Since the RIS-assisted system with a rank-one BS-to-RIS LoS channel can efficiently serve only a single user, so we will now focus on the optimization of the RIS phases in a single-user setting, for which $\tau^*$ can be obtained using (\ref{tau2}). Problem \textit{(P2)} is written as,
 \begin{subequations}
 \begin{alignat}{2} \textit{(P3)} \hspace{.35in}
&\!\max_{\boldsymbol{\Phi}}      &\qquad&  \textbf{h}_{2}^{H}\textbf{R}_{RIS}^{1/2}\boldsymbol{\Phi}^{H}\textbf{b}\textbf{b}^{H} \boldsymbol{\Phi} \textbf{R}_{RIS}^{1/2} \textbf{h}_{2} \label{P3}\\
&\text{subject to} &      & |\phi_{n}|=1, \hspace{.08in} n=1,\dots, N.\label{constraint4}
\end{alignat}
\end{subequations}
By applying the change of variables $\textbf{h}_{2}^{H}\textbf{R}_{RIS}^{1/2}\boldsymbol{\Phi}^{H}\textbf{b}=\alpha \textbf{v}^{H}\bar{\textbf{g}}$, where $\textbf{v}=[v_{1}, \dots, v_{N}]^{T}$, $v_{n}=\phi_{n}$, $\forall n$ and $\bar{\textbf{g}}=(\text{diag}(\textbf{h}_{2}^{H})\textbf{R}_{RIS}^{1/2})^{T}\textbf{b}$, \textit{(P3)} can be written equivalently as,
 \begin{subequations}
 \begin{alignat}{2} \textit{(P4)} \hspace{.35in}
&\!\max_{\textbf{v}}      &\qquad& \alpha^2 \textbf{v}^{H}\bar{\textbf{g}}\bar{\textbf{g}}^{H}\textbf{v} \label{P4}\\
&\text{subject to} &      & |v_{n}|=1, \hspace{.08in} n=1,\dots, N.\label{conn}
\end{alignat}
\end{subequations}
The optimal solution of \textit{(P4)} is,
\begin{align}
\label{sol} 
\textbf{v}^*=\exp(j \text{ arg}(\bar{\textbf{g}})).
\end{align}

\textit{Proof.} The problem can be solved optimally by noting that $\textbf{v}^{H}\bar{\textbf{g}}\bar{\textbf{g}}^{H}\textbf{v}=|\textbf{v}^{H}\bar{\textbf{g}}|^2$ We can write $|\textbf{v}^{H}\bar{\textbf{g}}|$ as $|\sum_{n=1}^N |v_n||\bar{g}_n|\exp(j (\text{arg}(\bar{g}_n)-\text{arg}(v_n))|$, where $\bar{g}_n$ is the $n^{th}$ element of $\bar{\textbf{g}}$. Now under the constraint in (\ref{conn}), we have $|v_{n}|=1$. Moreover $|\bar{g}_n|$ is independent of \textbf{v}. The maximum value $\alpha^2|\textbf{v}^{H}\bar{\textbf{g}}|^2$ can take under the constraint in  (\ref{conn}) will therefore be  $\alpha^2 (\sum_{n=1}^N |\bar{g}_n|)^2$, which is when $\text{arg}(\bar{g}_n)=\text{arg}(v_n)$, $\forall n$. Therefore the optimal solution is (\ref{sol}).
\vspace{-.1in}
\section{RIS Design for Full-Rank $\textbf{H}_{1}$}
In the last section, we showed that with a rank-one channel between the BS and the RIS, deploying an RIS yields no substantial benefit in a multi-user setting. To benefit from the RIS in a multi-user setting, rank($\textbf{H}_{1}$) must be greater than $K$. One way to introduce this rank is to have several RISs in the LoS of the BS as discussed at the end of Section III-A. Other methods are to introduce some deterministic scattering between the BS and the RIS or increase the spacing between the RIS elements. Moreover, by positioning the RIS close to the BS or by using large RISs, the BS-to-RIS LoS channel will be better represented by the spherical wave model and can have a high rank \cite{LoS1}. In this section, we will assume that $\textbf{H}_{1}$ is a full-rank LoS channel matrix and resort to the asymptotic analysis of the OLP using tools from RMT. Specifically, we will develop deterministic approximations for the minimum SINR under the OLP, $\tau^*$, given in (\ref{tau}), the quantities $q_{k}^*$s, given in (\ref{q}) and the optimal allocated powers $p_{k}^*$s, given in (\ref{P_opt}). These deterministic equivalents will be utilized to propose an algorithm to design the RIS parameters. 


To make the problem analytically more tractable, we consider a system with a common user channel correlation matrix \cite{Abla, Abla_common, Abla1}. Despite possible in principle, the extension to the case in which users have different channel correlation matrices is mathematically much more involved and is left for future work. In practice, the considered scenario may arise in networks where users are clustered on the basis of their covariance matrices such that users with similar covariance matrices are put in the same cluster \cite{Abla_common1, Abla_common2}.  We would highlight here that all existing works on RIS as well as OLP assume $\textbf{R}_{RIS_k}=\textbf{I}_N$, $k=1,\dots, K$ \cite{MMS, LIS, LIS1, LIS2,  LIS_new1, LIS_CS,RIS, RIS1, RIS2,mishra}.  We are the first to analyze the OLP under an arbitrary spatial correlation matrix at the RIS. 

\vspace{-.1in}
\subsection{Large System Analysis}
As shown in Section II, the OLP and $\tau^*$ are parametrized by the scalars $q^*_{k}$s that need to be evaluated by an iterative procedure using (\ref{q}) and (\ref{tau}). This is a computationally demanding task, especially when $M,N, K$ are large, since the matrix inversion operation in (\ref{q}) and (\ref{tau}) must be recomputed at every iteration. In addition, both depend directly on the channel vectors $\textbf{h}_{k}$ and change at the same pace as the small-scale fading (i.e., at the order of milliseconds). Therefore their computation needs to be performed at every channel realization. Moreover, computing $q_k$s as the fixed point of (\ref{q}) and (\ref{tau}) does not provide any insight into its optimal structure. To overcome these issues, we exploit the statistical distribution of $\textbf{h}_{k}$ and the large values of $M, N, K$ to compute deterministic approximations of $q^*_{k}$ and $\tau^*$. These approximations will only depend on the users' path losses and slowly varying spatial correlation matrices and will not have to be computed at each realization. Getting the deterministic approximation for the SINR $\tau^*$ under OLP will also allow us to optimize the RIS phases using information on only the channel large-scale statistics. For technical purposes, we shall consider the following assumptions:

\textbf{Assumption 3.} $M$, $N$ and $K$ grow large with a bounded ratio  as $ 0 < \text{lim inf}\frac{K}{M} \leq \text{lim sup} \frac{K}{M} < \infty$ and $0 < \text{lim inf}\frac{N}{M} \leq \text{lim sup} \frac{N}{M} < \infty$. In the sequel, this assumption is denoted as $\xrightarrow[n \rightarrow \infty]{}$.

\textbf{Assumption 4.} The correlation matrix $\textbf{R}_{RIS}$ will satisfy $\text{lim sup}_{N} ||\textbf{R}_{RIS}|| < \infty$.

Next, we resort to the large dimension analysis using RMT to show that $q_{k}^*$ and $\tau^*$ get asymptotically close to explicit deterministic quantities, as stated in the following theorem.

\textbf{Theorem 3.} Under the setting of \textbf{Assumptions 1, 3} and \textbf{4}, we have $|\tau^*-\bar{\tau}| \xrightarrow[n \rightarrow \infty]{a.s.} 0$, where $\bar{\tau}$ is given as the unique positive solution to the following fixed point equation:
\begin{align}
\label{tau_bar1}
&\bar{\tau}=\frac{1}{K} \text{tr }\textbf{R}\left(\frac{1}{1+\bar{\tau}}\textbf{R}+\xi \textbf{I}_{M} \right)^{-1} ,
\end{align}
where $\textbf{R}=\textbf{H}_{1}\boldsymbol{\Phi}\textbf{R}_{RIS} \boldsymbol{\Phi}^H \textbf{H}_1^H$ and $\xi=\frac{1}{P_{max}} \frac{1}{K}\sum_{j=1}^{K}\frac{1}{\rho_{j}}$.  Also, $\text{max}_{k}|q_{k}^*-\bar{q}_{k}| \xrightarrow[n \rightarrow \infty]{a.s.} 0$, where,
\begin{align}
\label{q_bar_c}
&\bar{q}_{k}=\frac{P_{max}}{\rho_{k}} \frac{1}{\frac{1}{K}\sum_{j=1}^{K}\frac{1}{\rho_{j}}}.
\end{align}

\textit{Proof:} The proof is detailed in Appendix C.

Note that these deterministic equivalents are tight for moderate system dimensions as well \cite{Abla, Abla_common, Abla1,SINRdeterministic, MMS, ourworkTCOM}. The above theorem provides an explicit form for $\bar{q}_{k}$s whose computation requires only the knowledge of the channel attenuation coefficients $\beta_{k}$s and $P_{max}$. Note that in the downlink the parameter $q_{k}$ is known to act as user $k$'s  priority parameter and in the uplink it corresponds to the transmit power of user $k$. Consequently, (\ref{q_bar_c}) indicates that in the asymptotic regime more power is given to users with weaker channel conditions. 

An important consequence of \textbf{Theorem 3} is that the performance of the network remains asymptotically the same if $q^*_{k}$ is replaced with $\bar{q}_{k}$, such that the asymptotically optimal linear precoding vector for user $k$ in (\ref{G_opt}) is computed as,
\begin{align}
\label{G_bar}
\bar{\textbf{g}}_{k}=\frac{\left(\sum_{i\neq k} \frac{\bar{q}_{i}}{K}\textbf{h}_{i}\textbf{h}_{i}^{H}+\textbf{I}_{M}\right)^{-1}\textbf{h}_{k}}{||\left(\sum_{i\neq k} \frac{\bar{q}_{i}}{K}\textbf{h}_{i}\textbf{h}_{i}^{H}+\textbf{I}_{M}\right)^{-1}\textbf{h}_{k}||}.
\end{align}

Next we provide the deterministic equivalent for the optimal transmit powers $p_{k}^*$. 

\textbf{Theorem 4.} Under the setting of \textbf{Theorem 3}, we have $\text{max}_{k}|p_{k}^*-\bar{p}_{k}| \xrightarrow[n \rightarrow \infty]{a.s.} 0$, where,
\begin{align}
\label{p_det}
&\bar{p}_{k}=\frac{\bar{\tau}}{\rho_{k}\zeta^2}\left(\frac{1}{(1+\bar{\tau})^2}P_{max}\rho_{k}\bar{\zeta} +\tilde{\zeta} \right), 
\end{align}
where $\zeta$ is given as the unique solution to,
\begin{align}
&\zeta=\frac{1}{M}\text{tr }\textbf{R}\bar{\textbf{T}}, \hspace{.1in} \tilde{\zeta}=\frac{\frac{1}{K}\text{tr }\textbf{R}\bar{\textbf{T}}\bar{\textbf{T}}}{1-\frac{\bar{\tau}^2}{\zeta^2(1+\bar{\tau})^2}\frac{1}{K}\text{tr }\textbf{R}\bar{\textbf{T}}\textbf{R}\bar{\textbf{T}}}, 
\end{align}
\begin{align}
&\bar{\zeta}=\frac{\frac{1}{K}\text{tr }\textbf{R}\bar{\textbf{T}}\textbf{R}\bar{\textbf{T}}}{1-\frac{\bar{\tau}^2}{\zeta^2(1+\bar{\tau})^2}\frac{1}{K}\text{tr }\textbf{R}\bar{\textbf{T}}\textbf{R}\bar{\textbf{T}}}, \hspace{.1in} \bar{\textbf{T}}=\left(\frac{\bar{\tau}\textbf{R}}{\zeta(1+\bar{\tau})}+ \textbf{I}_{M}  \right)^{-1}.
\end{align}

\textit{Proof:} The proof follows along the same arguments as those used for $q_{k}$s.

These results are particularly interesting from an implementation point of view. Indeed, unlike  $p^*_{k}$, $q^{*}_{k}$ and $\tau^*$, $\bar{p}_{k}$, $\bar{q}_{k}$ and $\bar{\tau}$ depend only on the large-scale channel statistics, including the slowly varying spatial correlation matrices, that can be estimated using knowledge of only the mean angles and angular spreads, and the channel attenuation coefficients, that change slowly with time. As a consequence, $\bar{p}_{k}$, $\bar{q}_{k}$s and $\bar{\tau}$ are not required to be computed at every channel realization. This provides a substantial reduction in the computational complexity as compared to the OLP since solving (\ref{P_opt}), (\ref{q}) and (\ref{tau}) at the pace of fast fading channel is no longer required.




\vspace{-.15in}
\subsection{Optimization of $\boldsymbol{\Phi}$}
\vspace{-.05in}

We will now solve problem \textit{(P2)} for large systems by replacing $\tau^*$ with its deterministic equivalent $\bar{\tau}$. The immediate benefit is that the designed algorithm will depend only  on the channel large scale statistics and therefore the elements of $\boldsymbol{\Phi}$ will not need to be optimized at every channel realization. This will result in significant reductions in the signal exchange overhead between the BS and RIS, as the BS will need to update $\boldsymbol{\Phi}$ and provide this information to RIS controller after several coherence intervals. Moreover the BS and RIS operations will need to be synchronized only when the correlation matrices change, reducing the real-time delays incurred each time RIS parameters are re-adjusted. For a large system, \textit{(P2)} is stated as:
\vspace{-.02in}
 \begin{subequations}
 \begin{alignat}{2} \textit{(P5)} \hspace{.35in}
&\!\max_{\boldsymbol{\Phi}}      &\qquad&  \bar{\tau}=\frac{1}{K} \text{tr }\textbf{H}_{1}\boldsymbol{\Phi}\textbf{R}_{RIS}\boldsymbol{\Phi}^{H}\textbf{H}_{1}^H \left(\frac{1}{1+\bar{\tau}}\textbf{H}_{1}\boldsymbol{\Phi}\textbf{R}_{RIS}\boldsymbol{\Phi}^{H}\textbf{H}_{1}^H+\xi \textbf{I}_{M} \right)^{-1} \label{P6}\\
&\text{subject to} &      & |\phi_{n}|=1, \hspace{.08in} n=1,\dots, N.\label{constraint7}
\end{alignat}
\end{subequations} \vspace{-.02in}
The optimization problem in \textit{(P5)} is a constrained maximization problem and can be solved using projected gradient ascent. We employ gradient search to monotonically increase the objective function of (\textit{P5}), eventually converging to a stationary point. At each step the solution is projected onto the closest feasible point that satisfies the constraint in (\ref{constraint7}). To proceed, we need the derivative of $\bar{\tau}$ with respect to each $\phi_{n}$, $n=1,\dots, N$ (recall that $\boldsymbol{\Phi}=\alpha \text{diag}(\phi_{1}, \dots, \phi_{N})$).

\textbf{Lemma 2.} The derivative of $\bar{\tau}$ defined in (\ref{P6})  with respect to $\phi_{n}$ can be computed as, \vspace{-.02in}
\begin{align}
\label{der}
&\frac{\partial \bar{\tau}}{\partial \phi_{n}}=\frac{2 \alpha}{K}\frac{\left[-\frac{1}{1+\bar{\tau}} \textbf{H}_1^H\textbf{T} \textbf{H}_1 \boldsymbol{\Phi} \textbf{R}_{RIS} \boldsymbol{\Phi}^H \textbf{H}_1^H \textbf{T} \textbf{H}_1 \boldsymbol{\Phi} \textbf{R}_{RIS} + \textbf{H}_1^H\textbf{T} \textbf{H}_1 \boldsymbol{\Phi} \textbf{R}_{RIS}  \right]_{n,n}}{1-\frac{1}{K} \frac{1}{(1+\bar{\tau})^2}\text{tr }\textbf{H}_{1}\boldsymbol{\Phi}\textbf{R}_{RIS}\boldsymbol{\Phi}^{H}\textbf{H}_{1}^H \textbf{T}\textbf{H}_{1}\boldsymbol{\Phi}\textbf{R}_{RIS}\boldsymbol{\Phi}^{H}\textbf{H}_{1}^H  \textbf{T}},
\end{align} \vspace{-.02in}
where $\textbf{T}=\left(\frac{1}{1+\bar{\tau}}\textbf{H}_{1}\boldsymbol{\Phi}\textbf{R}_{RIS}\boldsymbol{\Phi}^{H}\textbf{H}_{1}^H+\xi \textbf{I}_{M} \right)^{-1}$.

\textit{Proof:} The proof of Lemma 2 is provided in Appendix D. 

To proceed with this approach, a suitable step size $\mu$ for the gradient ascent needs to be computed at each iteration, for which we use the backtracking line search \cite{Boyd}. To this end, denote by $\textbf{v}^{k}=[\phi^{k}_{1}, \dots, \phi^{k}_{N}]^{T}$ the induced phases at step $k$ and by  $\textbf{p}^{k}$ the adopted ascent direction at step $k$, where $[\textbf{p}^{k}]_{n}=\frac{\partial \bar{\tau}}{\partial \phi_{n}}$. Then, the next iteration point is given by,
\begin{align}
&\tilde{\textbf{v}}^{k+1}=\textbf{v}^{k}+\mu \textbf{p}^{k}, \\
\label{impp}
&\textbf{v}^{k+1}=\exp(j \text{arg } (\tilde{\textbf{v}}^{k+1})).
\end{align}
The solution in (\ref{impp}) is obtained  by solving the projection problem $\min_{|v_{n}|=1, n=1,\dots, N} ||\textbf{v}-\tilde{\textbf{v}}||^2 $ to satisfy the constraint in (\ref{constraint7}). The complete algorithm is outlined in \textbf{Algorithm 1}. 

\begin{algorithm}[!t]
\caption{Projected Gradient Ascent Algorithm for the RIS Design}\label{alg:euclid}
\begin{algorithmic}[1]
\Procedure{Design of RIS Phases}{$\boldsymbol{\Phi}^*$}
\State \textbf{Initialize:} $\textbf{v}^o=\exp(j\pi/2)\textbf{1}_{N}$, $\boldsymbol{\Phi}^{o}=\text{diag}(\textbf{v}^o)$, $\bar{\tau}^o=f(\bar{\tau}^o, \boldsymbol{\Phi}^{o})$ given by (\ref{P6}), $\epsilon>0$;
\State  \textbf{for} $k=0, 1, 2, \dots$, do
\State $[\textbf{p}^{k}]_{n}=\frac{\partial \bar{\tau}^k}{\partial \phi_{n}}$, $n=1,\dots, N$, where $\frac{\partial \bar{\tau}}{\partial \phi_{n}}$ is given in \textbf{Lemma 2};
\State $\mu$=\text{backtrack line search}($f(\bar{\tau}^k, \boldsymbol{\Phi}^{k})$, $\textbf{p}^k$, $\textbf{v}^{k}$) \cite{Boyd};
\State $\tilde{\textbf{v}}^{k+1}=\textbf{v}^{k}+\mu \textbf{p}^{k}$;
\State $\textbf{v}^{k+1}=\exp(j \text{arg } (\tilde{\textbf{v}}^{k+1}))$;
\State $\boldsymbol{\Phi}^{k+1}=\alpha \text{diag}(\textbf{v}^{k+1})$;
\State $\bar{\tau}^{k+1}=f(\bar{\tau}^{k+1}, \boldsymbol{\Phi}^{k+1})$ ;
\State \textbf{Until} $||\bar{\tau}^{k+1}-\bar{\tau}^{k}||^2< \epsilon$; Obtain $\boldsymbol{\Phi}^*=\boldsymbol{\Phi}^{k+1}$;
\State \textbf{end for}
\EndProcedure
\State $\boldsymbol{\Phi}^*$ fed to the RIS controller by the BS;
\State BS computes the asymptotic OLP $\bar{\textbf{G}}^*=[\bar{\textbf{g}}_{1}^*, \dots, \bar{\textbf{g}}_{K}^*]$ using (\ref{G_bar}) and $\bar{\textbf{p}}^*$ using (\ref{p_det});
\end{algorithmic}
\end{algorithm}

At each step, the proposed algorithm increases the value of the deterministic equivalent of the minimum SINR, thus converging in the value of the objective. However, no global optimality claim can be made due to the fact that the problem \textit{(P5)} is not convex with respect to $\boldsymbol{\Phi}$. Since this a preliminary work on optimizing the max-min SINR performance of a RIS-assisted multi-user MISO system, so this algorithm is a good starting point to study this new technology. Note that the algorithm does not require the computation of the optimized phases at every channel realization but only once over several coherence intervals, providing a substantial reduction in computational complexity and signal exchange overhead between the BS and the RIS controller.

\vspace{-.15in}
\subsection{Results with Direct Channel - Special Case}

In scenarios, where the direct channels between the BS and the users, distributed as $\textbf{h}_{d,k} \sim \mathcal{CN}(\textbf{0},\textbf{I}_M)$, are also available, the overall channel is given as,
\begin{align}
\label{ch_d}
&\textbf{h}_k=\sqrt{\rho_k} \textbf{H}_1 \boldsymbol{\Phi} \textbf{R}_{RIS}^{1/2} \textbf{h}_{2,k}+\sqrt{\rho_{d,k}} \textbf{h}_{d,k},
\end{align}
where $\rho_{d,k}$ is the average SNR of direct link. The RMT analysis done in this section can not be easily extended  to this scenario. To see this note that in our proposed method in Appendix C to derive the deterministic equivalents of $q_k^*$ and $\tau^*$, we introduce the quantities $d_k=\frac{1}{K} \tilde{\textbf{h}}_k^H \textbf{Q}_k \tilde{\textbf{h}}_k$, $k=1,\dots, K$, where $\tilde{\textbf{h}}_k=\rho_k^{-1/2} \textbf{h}_k$ and therefore $\tilde{\textbf{h}}_k\sim \mathcal{CN}(\textbf{0}, \textbf{R})$, where $\textbf{R}=\textbf{H}_1\boldsymbol{\Phi}\textbf{R}_{RIS} \boldsymbol{\Phi}^H \textbf{H}_1^H$  is the same for all users. This implies that using RMT, all $d_k$s should converge to the same quantity $\tilde{d}$. A guess was then made for $\tilde{d}$ and it was rigorously proved that the guess is correct i.e. the condition in (\ref{cond}) was verified. Now if the direct channel is included,  there is no way to define $\tilde{\textbf{h}}_k$ such that the correlation matrix characterizing $\tilde{\textbf{h}}_k$ is independent of $k$. Therefore each $d_k$ will converge to a different $\tilde{d}_{k}$. We can make a guess for $\tilde{d}_k$ based on similar arguments as done in Appendix C. However, so far, there is no available method using RMT tools to rigorously prove that $\text{max}_{k}|\frac{d_k}{\tilde{d}_k}-1 |\rightarrow 0$ for such a setting.  Therefore the OLP has never been analyzed for scenarios with per-user correlation matrices or where the overall channel is the sum of two channels with different path losses. This is why for the main body of this work, we focused our attention on scenarios where the direct link are easily blocked as done in \cite{LIS1, LIS2, LIS_new1, RIS, RIS1, RIS2}. 

However, there is a special case where our analysis can be extended, which is where $\frac{\rho_{d,k}}{\rho_{k}}=c$, i.e. the ratio of the average SNR of direct link to the average SNR of RIS-assisted link is the same for each user. This condition implies that $\frac{\beta_{d,k}}{\beta_{k}}=c$, where $c$ is independent of $k$.  Such a condition can  be met when the length of RIS-user link and BS-user link are approximately equal, which may happen in scenarios where RIS is located very close to the BS or the users are located far from the BS and RIS. For such a setting, we can define $\tilde{\textbf{h}}_k=\rho_k^{-1/2} \textbf{h}_k$ and therefore $\tilde{\textbf{h}}_k \sim \mathcal{CN}(\textbf{0}, \tilde{\textbf{R}})$, where $\tilde{\textbf{R}}=\textbf{R}+c\textbf{I}_M$. Under this case, \textbf{Theorem 3} is given as \textbf{Corollary 3}.

\textbf{Corollary 3.} Under the setting of \textbf{Assumptions 1, 3} and \textbf{4} and the channel in (\ref{ch_d}), \textbf{Theorem 3} is given as,
\begin{align}
\label{tau_bar11}
&\bar{\tau}=\frac{1}{K} \text{tr }\tilde{\textbf{R}}\left(\frac{1}{1+\bar{\tau}}\tilde{\textbf{R}}+\xi \textbf{I}_{M} \right)^{-1} , \hspace{.1in} \bar{q}_{k}=\frac{P_{max}}{\rho_{k}} \frac{1}{\frac{1}{K}\sum_{j=1}^{K}\frac{1}{\rho_{j}}}.
\end{align}
where $\tilde{\textbf{R}}=\textbf{H}_{1}\boldsymbol{\Phi}\textbf{R}_{RIS} \boldsymbol{\Phi}^H \textbf{H}_1^H+c \textbf{I}_{M}$ and $\xi=\frac{1}{P_{max}} \frac{1}{K}\sum_{j=1}^{K}\frac{1}{\rho_{j}}$. Moreover, \textbf{Algorithm 1} can be used to design the RIS phases with $\frac{\partial \bar{\tau}}{\partial \phi_{n}}$ given by (\ref{der}) with \textbf{R} replaced with $\tilde{\textbf{R}}$.
%

\section{Simulation Results and Discussions}

In the simulations, we consider a uniform linear array of antennas and reflecting elements at the BS and the RIS respectively. The coordinates of BS, RIS and user are denoted as $(x_{BS},y_{BS},25)$, $(x_{RIS},y_{RIS},40)$ and $(x_{U}, y_{U},1.5)$ respectively. The generation of the correlation matrix for the RIS largely depends on the underlying technology being used to realize this surface. So far, there has not been much effort in characterizing the spatial correlation for RISs realized using reconfigurable meta-surfaces. However, the channels for passive reflect-arrays are generated using the phased array model utilized for conventional antenna arrays.  Assuming that RIS is realized using a passive reflect-array \cite{RA, reflective}, we utilize the correlation model developed for the conventional linear antenna array in \cite{ourwork, correlation4}.  Specifically, the 3D correlation coefficient between the RIS elements $n$ and $n'$, with respect to user $k$, is given as \cite{ourwork},
\begin{align}
\label{corr}
&[\textbf{R}_{RIS_k}]_{n,n'}=\mathbb{E}\left[\exp\left(j\frac{2\pi}{\lambda}d_{RIS}(n-n')\sin\phi_{k} \sin\theta_{k} \right)\right],
\end{align}
where $d_{RIS}$ is the inter-element separation, $\theta_k$ represents the elevation angles and $\phi_k$ represents the azimuth angles for user $k$. The elevation angles are generated using the Laplace distribution with mean angle of departure (AoD), $\theta_{0}(k)$, and spread $\sigma^2_{l}=8^o$ and the azimuth angles are generated using the Von Mises distribution with mean AoD $\mu(k)$ and spread $\frac{1}{\kappa}$, with $\kappa=5$ \cite{ourwork}. The channel attenuation coefficients $\beta_{k}=\beta_1 \beta_{2,k}$, where $\beta_1$ and $\beta_{2,k}$ are the channel attenuation coefficients for the BS-to-RIS link and RIS-to-user $k$ link respectively. We model them using the 3GPP Urban Micro (UMi) parameters from [\cite{TR36.814}, Table B.1.2.1-1] at $2.5$ GHz operating frequency. We use the LoS version to generate $\beta_1$ and the non-LOS (NLOS) version of the UMi scenario to generate $\beta_{2,k}$s. We let $G_t$ and $G_r$ denote the antenna gains (in \rm{dBi}) at the transmitter and receiver respectively and assume that the elements of BS and RIS have $5\rm{dBi}$  gain while each user is quipped with a single $0 \rm{dBi}$ antenna \cite{relay_p2, LIS}. The expressions of $\beta_1$ and $\beta_{2,k}$ are,
\begin{align}
&\beta_1=G_t+G_r-35.95-22\log_{10} d_{BS-RIS}, \hspace{.1in} \beta_{2,k}=G_t+G_r-33.05-36.7 \log_{10} d_{RIS-Uk} \nonumber,
\end{align}
where $d_{BS-RIS}$ and $d_{RIS-Uk}$ are the distances between BS and RIS and RIS and user $k$ respectively. We set $P_{max}=5W$, the bandwidth $B=180kHz$ and $\sigma^2=-174+10 \log_{10} B$ in $\rm{dBm}$.

\begin{figure}[!t]
\begin{minipage}[b]{0.45\linewidth}
\centering
\includegraphics[width=2.5 in]{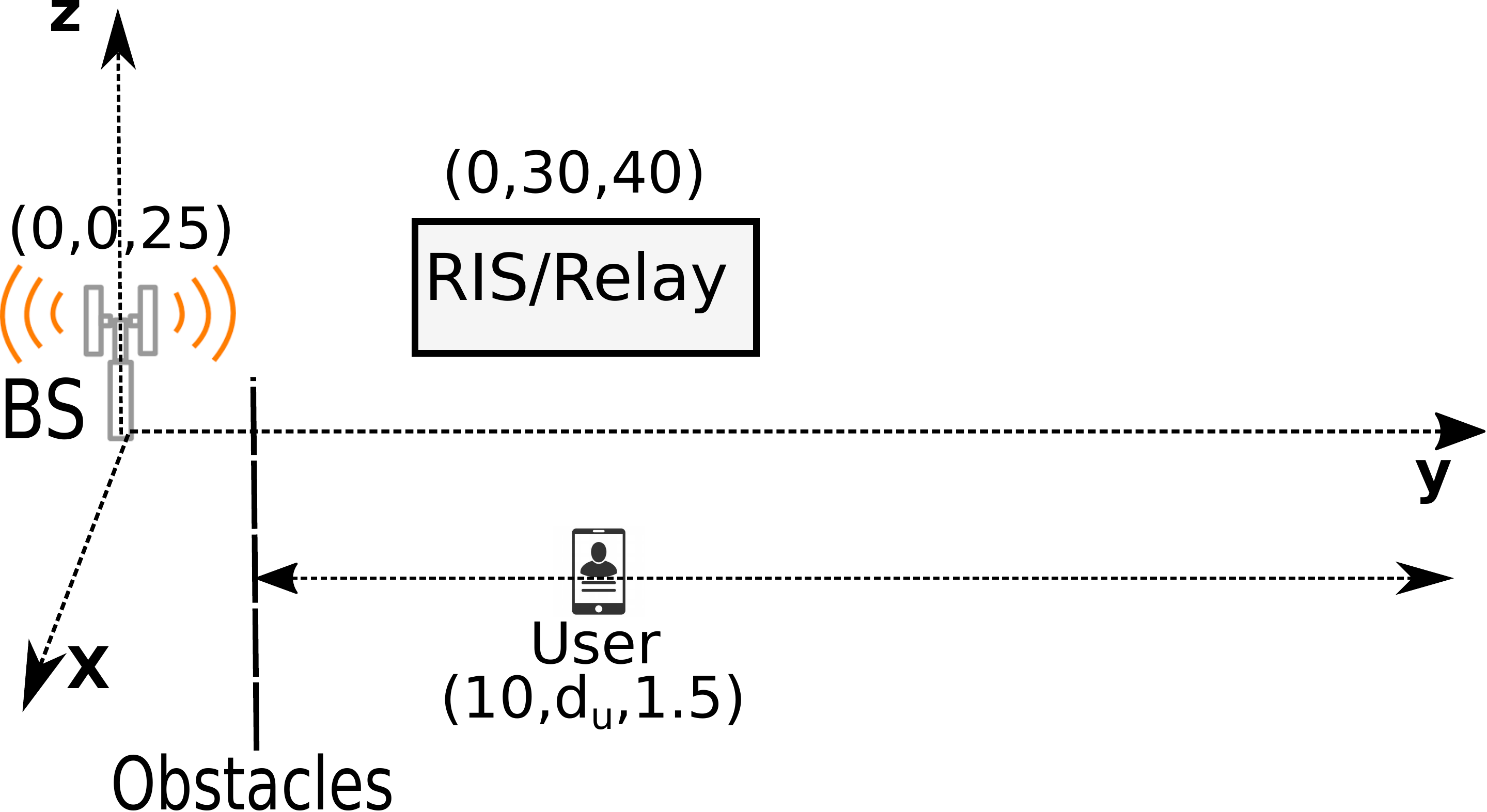}
\caption{Single user layout.}
\label{singleuserlayout}
\end{minipage}
\hspace{0.5cm}
\begin{minipage}[b]{0.45\linewidth}
\centering
\includegraphics[width=2.85in, height=2in]{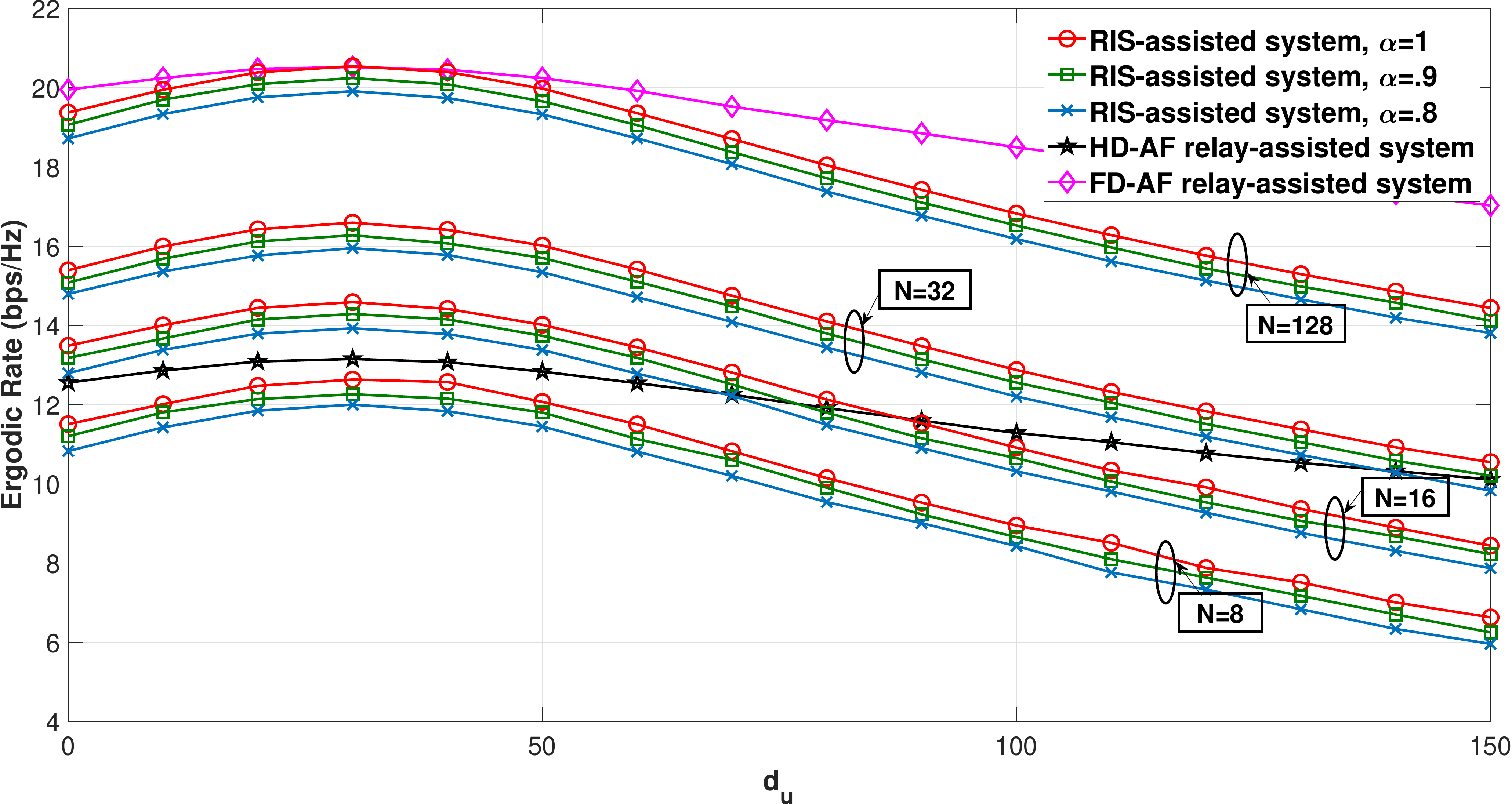}
\caption{Performance comparison of the RIS with HD-AF and FD-AF relays in a single-user MISO system.}
\label{single1}
\end{minipage}
\end{figure}
\vspace{-.1in}
\subsection{Rank-One $\textbf{H}_{1}$}
We first focus on the scenario when $\textbf{H}_{1}$ is a rank-one matrix, modeled as $\textbf{H}_{1}=\textbf{a}\textbf{b}^H$, where $\textbf{a}_{m}=\exp\left(j\frac{2\pi}{\lambda}d_{BS}(m-1)\sin\phi_{LoS_1} \sin\theta_{LoS_1} \right)$ and $\textbf{b}_{n}=\exp\left(j\frac{2\pi}{\lambda}d_{RIS}(n-1)\sin\phi_{LoS_2} \sin\theta_{LoS_2} \right)$, where $d_{BS}$ is the inter-antenna separation at the BS, $\phi_{LoS_1}$ and $\theta_{LoS_1}$ represent the LoS azimuth and elevation AoDs at the BS and $\phi_{LoS_2}$ and $\theta_{LoS_2}$ represent the LoS azimuth and elevation angles of arrival (AoA)s at the RIS respectively. Note that in the simulations, we set $d_{BS}=d_{RIS}=0.5\lambda$.

\subsubsection{Single-User}
We first consider a single-user system, where the RIS is placed adjacent to the BS with $y_{RIS}=30m$. The system along with the coordinates has been shown in Fig. \ref{singleuserlayout}.  We study the received SNR of a user by varying the value of $d_{u}$. The optimized matrix $\boldsymbol{\Phi}^*$ is obtained using (\ref{sol}), where $\boldsymbol{\Phi}^*=\alpha \text{diag}(\textbf{v}^*)$ and the optimal precoder is MRT in (\ref{MRTT}).



The performance is compared to that of AF relay-assisted communication operating in HD and FD modes with $R$ antennas at the relay. For FD mode, we assume half of the relay antennas are used for transmission and the other half for reception. The power budgets at the BS and relay are $P_S$ and $P_R$ respectively. The user's ergodic rate expressions are given as \cite{relay_p},
\begin{align}
\label{HD}
&R_{HD-AF}=\frac{1}{2} \log_2 \left(1+\frac{\gamma_R^{HD} \gamma_D^{HD}}{\gamma_R^{HD}+\gamma_{D}^{HD}+1}  \right), \\
\label{FD}
&R_{FD-AF}= \log_2 \left(1+\frac{\gamma_R^{FD} \gamma_D^{FD}}{\gamma_R^{FD}+\gamma_{D}^{FD}+1}  \right),
\end{align}
where $\gamma_R^{HD}=\frac{P_S^* \beta_1 ||\textbf{H}_1||^2}{\sigma^2}$ and $\gamma_D^{HD}=\frac{P_R^* \beta_2 ||\textbf{R}_{RIS}^{1/2} \textbf{h}_{2}||^2}{\sigma^2}$. Note that $P_S^*, P_R^*$ are the optimal values that maximize $R_{HD-AF}$ such that $P_R^*+P_S^*=P_{max}$, and are given by [\cite{relay_p}, equation (40)]. Similarly, $\gamma_R^{FD}=\frac{P_S^* \beta_1 ||\tilde{\textbf{H}}_1||^2}{P_R \big|\frac{\tilde{\textbf{h}}_{2}^H}{||\tilde{\textbf{h}}_2||} \tilde{\textbf{h}}_{SI}  \big|^2 +\sigma^2}$ and $\gamma_D^{FD}=\frac{P_R^* \beta_2 ||\tilde{\textbf{R}}_{RIS}^{1/2} \tilde{\textbf{h}}_{2}||^2}{\sigma^2}$, where $\tilde{\textbf{H}}_1$ and $\tilde{\textbf{h}}_{2}$ are now $M \times \frac{R}{2}$ LoS BS-to-relay channel matrix and $\frac{R}{2}\times 1$ relay-to-user channel vector respectively. Moreover,  $\tilde{\textbf{h}}_{SI} \in \mathbb{C}^{\frac{R}{2}\times 1}$ denotes the loop-back  self-interference (SI) of the relay when it operates in FD mode. The elements of this vector are drawn from Rician distribution with mean equal to noise power (a very optimistic value). In practice, the amount of SI is higher depending on the efficiency of SI cancellation circuits. The values of $P_S^*, P_R^*$ are computed to maximize $R_{FD-AF}$ and are given by [\cite{relay_p}, equation (47)].  To ensure a fair comparison, we have considered $5\rm{dBi}$ relay antennas. 

The first result is shown in Fig. \ref{single1} for $M=8$, where we plot the receive SNR at the user with respect to $d_{u}$ for different number of elements at the RIS, i.e. $N=\{8, 16, 32, 128 \}$.  The performance is compared to that of HD-AF and FD-AF relays having $8$ antennas.  As compared to the relays, the performance of RIS decreases faster as the distance between the user and RIS increases. This is evident from the rate expressions of the relays in (\ref{HD}) and (\ref{FD}), where an increase in relay-to-user distance decreases both the numerator and denominator of the SNR. We also observe that a passive RIS with $32$ reflecting elements can outperform the HD-AF relay for the entire range of $d_u$. Moreover, a decrease in the amplitude reflection coefficient $\alpha$ from $1$ to $0.8$ decreases the achievable user rate but the decrease is not very significant. Note that there exists designs for reconfigurable meta-surfaces that are capable of achieving higher than $\alpha=.9$ reflection coefficients. The performance of FD-AF relay is obviously much better than HD-AF relay since it does not suffer from pre-log penalty. Note that the amount of SI experienced by the FD-AF relay is assumed to be very low in the simulations, resulting in its high performance. In practical systems, it will likely suffer from higher SI, resulting in a decrease in its performance. To achieve a performance comparable to that of FD-AF relay, the RIS would require a much higher number of elements. Finally, we observe that doubling $N$ in an RIS-assisted system results in a $2$ bit rate improvement, which implies that the SNR scales in the order of $N^2$. 

\begin{figure}[!t]
\begin{minipage}[b]{0.45\linewidth}
\centering
\includegraphics[width=2.85in, height=1.8in]{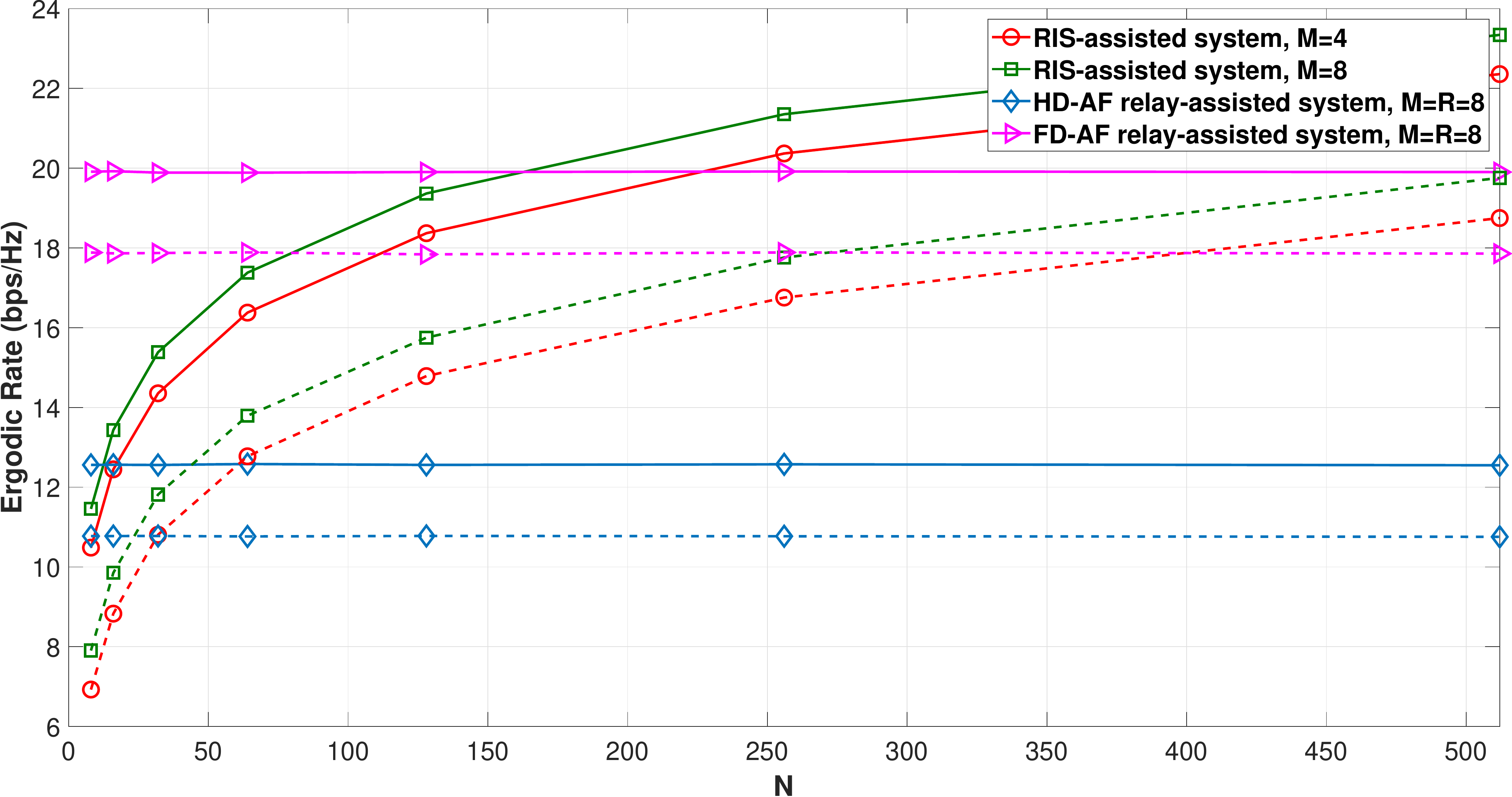}
\caption{RIS size needed to outperform HD-AF and FD-AF relays. Solid lines represent a user at $d_u=60m$ and dottled lines are for a user at $d_u=120m$.}
\label{with_N_single}
\end{minipage}
\hspace{0.5cm}
\begin{minipage}[b]{0.45\linewidth}
\centering
\includegraphics[width=2.5 in]{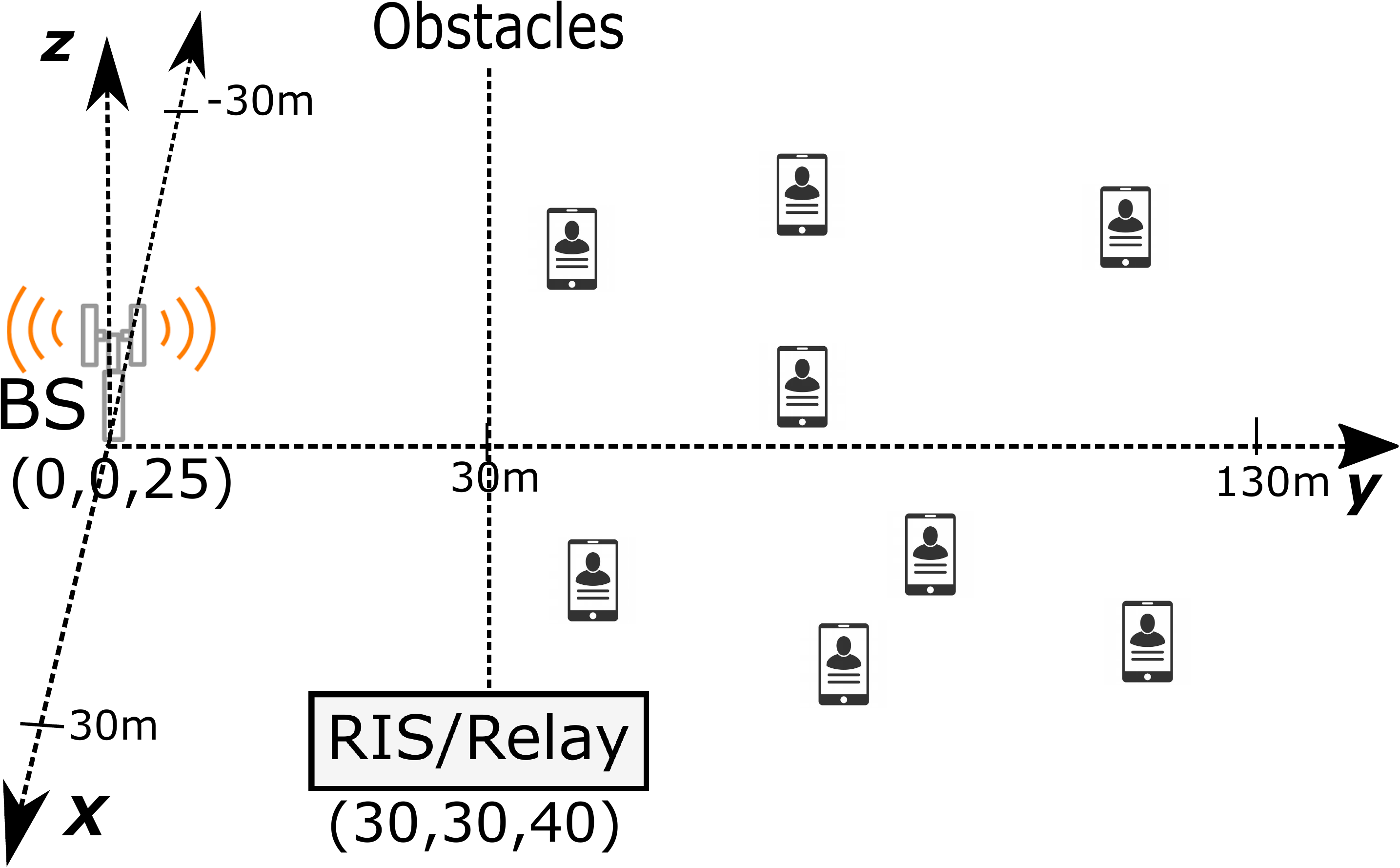}
\caption{Multi-user layout.}
\label{multiuserlayout}
\end{minipage}
\end{figure}

The next result in Fig. \ref{with_N_single} brings out the design criteria for selecting the size  $N$ for the RIS array to achieve the same performance as HD-AF and FD-AF relay-assisted systems with $8$ antennas at the BS and relay. The curves in solid lines are plotted for a user located at $d_u=60m$, where it is noted that an RIS-assisted system with $M = 8$ antennas at the BS and $N = 12$ reflecting elements at the RIS can achieve the same rate performance as that of the HD-AF relay whereas it needs $N=160$ element to achieve the performance of FD-AF relay. To achieve the  performance of HD-AF and FD-AF relay-assisted systems (with $M=8, R=8$) using an RIS-assisted system with only $M = 4$ antennas at the BS, the number of elements required at the RIS are $N = 18$ and $N=225$, respectively.  Therefore, an RIS-assisted system can realize the gains yielded by large MIMO arrays and relays with a lower number of active antennas at the BS, by using a large number of low-cost passive reflecting elements at the RIS. When the user moves to a distance of $120$m from the BS, the performance of both RIS-assisted and relay-assisted systems decreases, with the decrease being larger for RIS-assisted system as discussed earlier. A significantly larger number of elements is needed at the RIS to yield the same performance as relay-assisted systems. Therefore, RISs with moderate values of $N$ are competitive in small-cell scenarios, whereas large RISs are needed to outperform relays for cell-edge users in larger cells.

However, it is important to stress here that there are multiple reasons to consider the use of RISs instead of relays in future deployments \cite{relay_p1}. Relays are active devices that need a dedicated power source and active electronic components, such as digital-to-analog convertors (DACs), analog-to-digital converters (ADCs), mixers, power amplifiers for transmission, and low-noise amplifiers for reception.  The deployment of relays is, therefore, both costly and power-consuming. The hardware complexity of relays is further increased if FD relays are used due to the need for SI cancellation circuits. In contrast, RISs are meant to be realized with minimal hardware complexity without requiring dedicated power amplifiers, mixers, and DACs/ADCs. RISs are composite material layers made of metallic or dielectric patches printed on a grounded dielectric substrate. Their reconfigurability is ensured through low-power electronics (for example: switches) \cite{Renzo}. In fact, RISs are aimed to be of much lower cost and complexity than relay stations, especially at mass production as discussed in \cite{relay_p1}. An example of a large RIS made of inexpensive antennas can be found in \cite{ex_RIS}. RIS suffers from neither additive noise nor SI. Moreover,the average user SNR increases linearly with $R$, whereas it increases quadratically with $N$.  All these factors make RIS a more energy-efficient, low-cost and low-complexity solution than relays. 



%

\begin{figure}[!t]
\begin{minipage}[b]{0.45\linewidth}
\centering
\includegraphics[width=2.85in, height=1.9in]{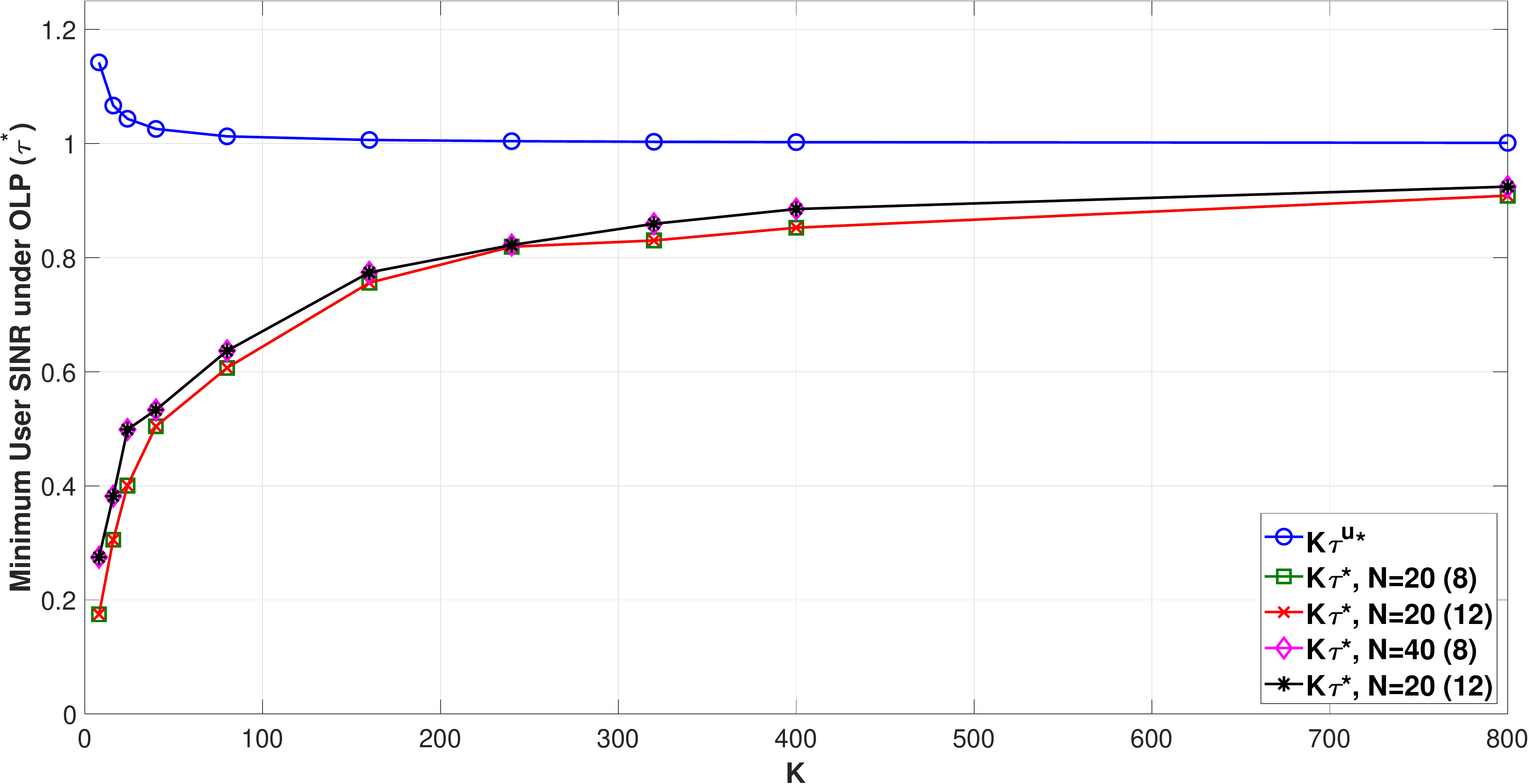}
\caption{Performance of the RIS in a multi-user MISO system under rank-one $\textbf{H}_{1}$.}
\label{multiuser_rankone}
\end{minipage}
\hspace{0.5cm}
\begin{minipage}[b]{0.45\linewidth}
\centering
\includegraphics[width=2.85in, height=1.9in]{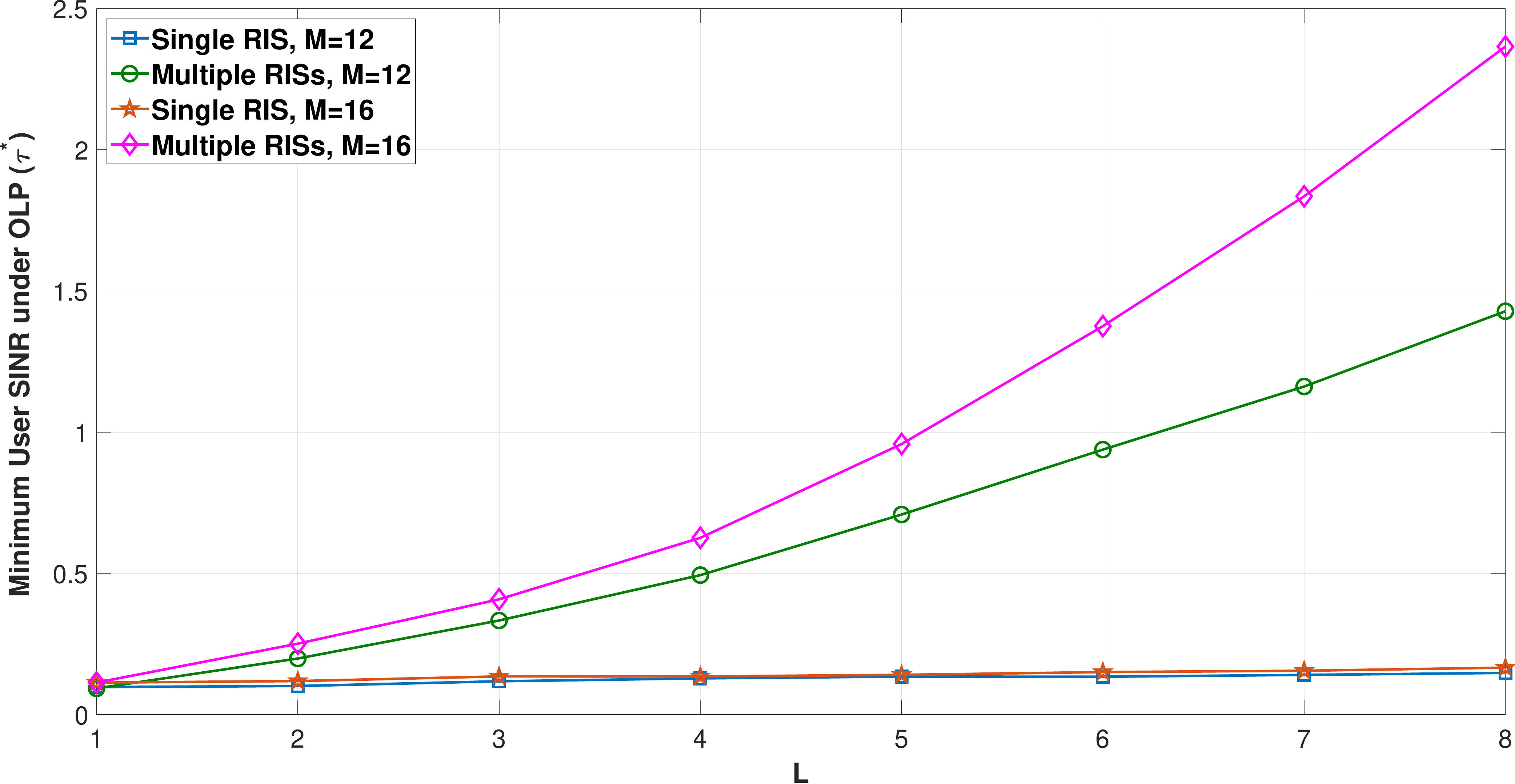}
\caption{Performance of a multi-user MISO system with multiple RISs under rank-one BS-to-RIS channels.}
\label{multiusermultiRIS}
\end{minipage}
\end{figure}

\subsubsection{Multiple Users}

We now study a system where the BS serves multiple users using an RIS that has a rank-one channel with the BS.  The BS, RIS and users locations are shown in Fig. \ref{multiuserlayout}. The phase matrix $\boldsymbol{\Phi}$ is set as $\boldsymbol{\Phi}(n,n)=\exp\left(j \frac{2\pi}{\lambda}(n-1)d_{RIS}\sin(\bar{\theta_0})\sin(\bar{\mu})\right)$, where $\bar{x}$ denotes the mean of $(x(1), \dots, x(K))$, i.e. the signals are reflected in the mean angular direction of the users. We refer to this scheme as Center of Means (CoM) phase adjustment.

The result in Fig. \ref{multiuser_rankone} plots the minimum user SINR $\tau^*$ under the OLP in (\ref{tau}) and its closed-form expression  given by (\ref{tau2}). Both results are plotted after scaling with $K$. We also plot the bound $K\tau^{*u}$, where $\tau^{*u}$ is given by (\ref{eq1}). In the result, we let $M,K$ grow at the same rate and observe that $K\tau^{*u}$ converges to $1$ with $K$ no matter how many RIS elements or BS antennas we use, thereby validating \textbf{Theorem 2}.  The exact values of the scaled minimum SINR stay below this bound, thereby showing that the system can not support multiple users under a rank-one link. 

Next we study in Fig. \ref{multiusermultiRIS} a system with multiple RISs, each with a rank-one BS-to-RIS channel. The minimum SINR under OLP is plotted for $L=8$ RISs deployed with $y_{RIS}=30$ and $x_{RIS}=\{40, 30, 20, 10, 0, -10, -20, -30\}$ and the performance is compared with that of a single RIS system for $K=5$ and $M=8$. As expected, the single-RIS system does not perform well, while, the minimum SINR increases as more RISs are introduced into the system. Note that for $L\leq 4$ the channel does not have enough rank to support $5$ users and therefore increasing $M$ does not make much of a difference. However, for $L\geq 5$, increasing $M$ would scale the performance due to the `massive MIMO effect' \cite{Abla1}. Therefore, in scenarios where each RIS is deployed to have a rank-one LoS channel with the BS, $L\geq K$ surfaces can be used to efficiently serve $K$ users, confirming the insights drawn from \textbf{Corollary 2}.

\vspace{-.15in}
\subsection{Full Rank $\textbf{H}_{1}$}
Next we assume a full rank BS-to-RIS LoS channel matrix given as $[\textbf{H}_{1}]_{m,n}=\exp( j\frac{2\pi}{\lambda}$
\begin{align}
&\times((m-1) d_{BS}\sin(\theta_{LoS_1}(n)) \sin(\phi_{LoS_1}(n))+(n-1) d_{RIS}\sin(\theta_{LoS_2}(n)) \sin(\phi_{LoS_2}(n))) ),
\end{align}
where $\theta_{LoS_1}(n)$ and $\phi_{LoS_1}(n)$, $n=1,\dots, N$, are generated uniformly between $0$ to $\pi$ and $0$ to $2\pi$ respectively. Note that $\theta_{LoS_2}(n)=\pi-\theta_{LoS_1}(n)$ and $\phi_{LoS_2}(n)=\pi+\phi_{LoS_1}(n)$. 

\begin{figure}[!t]
\begin{minipage}[b]{0.45\linewidth}
\centering
\includegraphics[width=2.85in, height=2in]{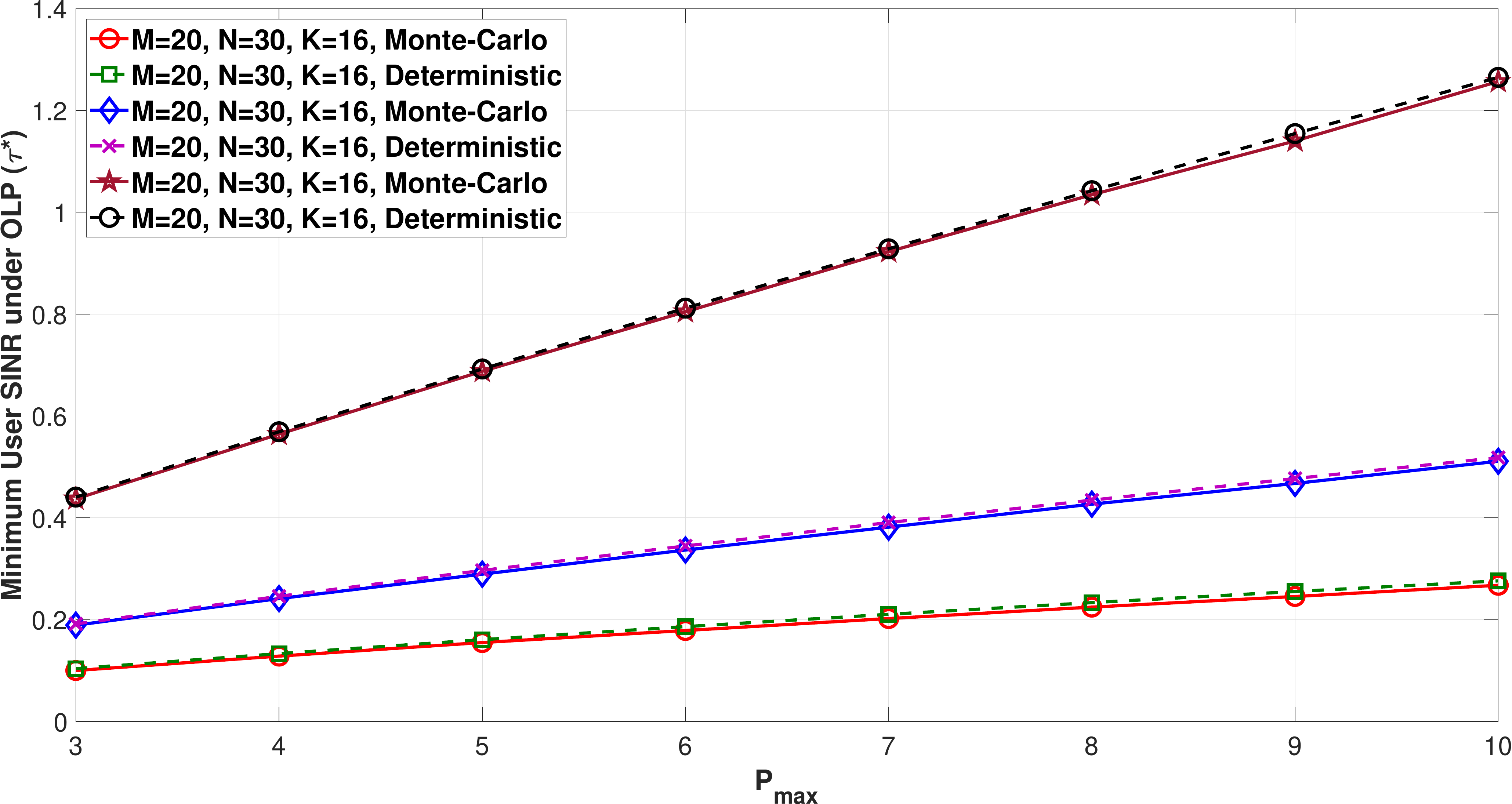}
\caption{Validation of the deterministic equivalent $\bar{\tau}$ for $\tau^*$ under full-rank $\textbf{H}_{1}$.}
\label{multiuser_fullrank}
\end{minipage}
\hspace{0.5cm}
\begin{minipage}[b]{0.45\linewidth}
\centering
\includegraphics[width=2.85in, height=2in]{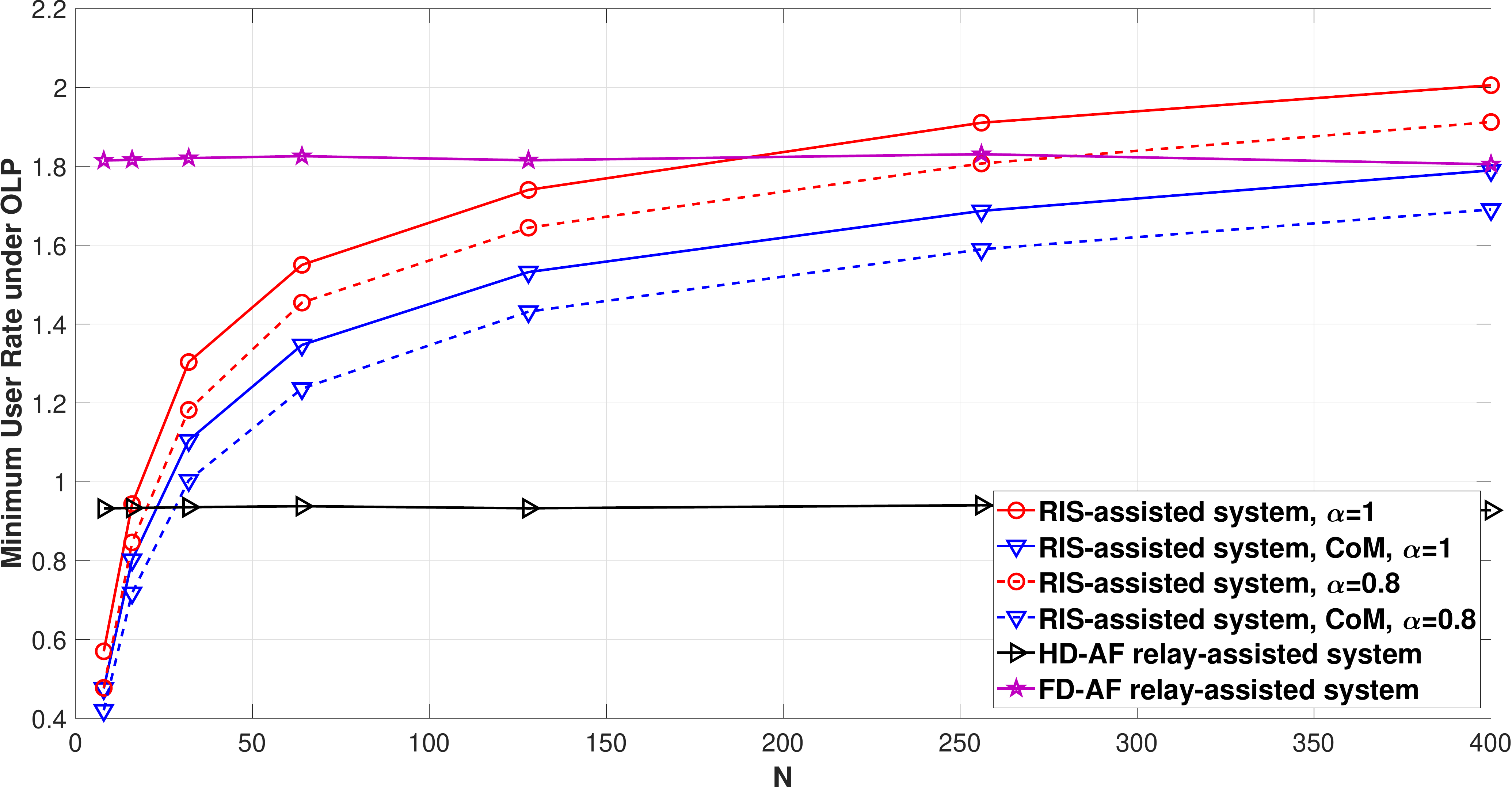}
\caption{RIS size needed under \textbf{Algorithm 1} to outperform FD-AF and HD-AF relays.}
\label{multiuser_fullrank_opt}
\end{minipage}
\end{figure}

We first check the asymptotic result in \textbf{Theorem 3} under the assumption of a common correlation matrix for all users, which is generated by using $\bar{\theta}_{0}$ and $\bar{\mu}$ in (\ref{corr}). We consider the multi-user setup in Fig. \ref{multiuserlayout} with CoM phase adjustment and plot both the Monte-Carlo simulated $\tau^*$ and the deterministic equivalent $\bar{\tau}$ against $P_{max}$ in Fig. \ref{multiuser_fullrank}. The deterministic equivalent provides a close match to the Monte-Carlo simulated minimum SINR even for moderate system dimensions. 

Next, we optimize the phase matrix $\boldsymbol{\Phi}$ using  \textbf{Algorithm 1} and study the minimum user rate results in Fig. \ref{multiuser_fullrank_opt} for different values of reflection coefficient $\alpha$ and $M,K=8$.  The minimum user rate for RIS-assisted system is computed as $\log_{2}(1+\bar{\tau})$, where $\bar{\tau}$ is given by (\ref{tau_bar1}). As compared to CoM reflect beamforming scheme, our proposed design performs significantly better. Moreover, a decrease in $\alpha$ results in a performance loss and therefore the availability of lossless RISs is important in considering the use of RISs in conventional MISO systems. 

The performance of RIS-assisted system is compared to the benchmark schemes of  HD-AF and FD-AF relays. The relay is assumed to be have $R=8$ antennas and is deployed at the same position as the RIS and therefore experiences same channels. In the considered AF relay scheme, the received signals at the relay and the user $k$ are given as,
\begin{align}
&\textbf{y}_R=\sqrt{\beta_1}\textbf{H}_1^H \textbf{x} + \textbf{n}_R, \hspace{.1in} y_k=\sqrt{\beta_{2,k}} \textbf{h}_{2,k}^H \textbf{R}_{\text{relay}}^{1/2}\textbf{V}^H \textbf{y}_R + n_k,
\end{align}
where $\textbf{V}$ is the diagonal AF matrix of the relay. The power constraint at the BS is $\mathbb{E}[|\textbf{x}|^2]=P_S$ and $\textbf{n}_R \sim \mathcal{CN}(\textbf{0},\sigma^2 \textbf{I}_R)$. The SINR at user $k$ can be expressed as $\frac{\frac{p_k}{K} |\textbf{h}_k^H \textbf{g}_k|^2}{\sum_{i\neq k} \frac{p_i}{K} |\textbf{h}_k^H \textbf{g}_i|^2+1}$, where $\textbf{h}_k=\sqrt{\rho_k} \textbf{H}_1 \textbf{V} \textbf{R}_{\text{relay}}^{1/2} \textbf{h}_{2,k}$ where $\rho_k=\frac{\beta_1 \beta_{2,k}}{\sigma^2 (1+\beta_{2,k} \textbf{h}_{2,k}^H \textbf{R}_{\text{relay}}^{1/2} \textbf{V}^H \textbf{V} \textbf{R}_{\text{relay}}^{1/2} \textbf{h}_{2,k})}$ and $\textbf{R}_{\text{relay}}$ is the correlation matrix at relay. Using the expression of $\textbf{h}_k$, the optimal precoding vectors $\textbf{g}_k$s and optimal powers $\textbf{p}_k^*$s are obtained as solution of (\ref{G_opt})-(\ref{P_opt}). Note that diagonal elements of $\textbf{V}$ are set as $\frac{\sqrt{P_R}}{\sqrt{\beta_1 P_{S} \textbf{h}_{1,n}^H \textbf{h}_{1,n}+\sigma^2}}$, $n=1,\dots, N$, to satisfy the power constraint on the relay as $\mathbb{E}[\text{tr}\textbf{V}^H \textbf{y}_R\textbf{y}_R^H \textbf{V}]=P_R$, where $P_S+P_R=P_{max}$. Note that the optimal split of $P_S$ and $P_R$ is found through numerical exhaustive search. The optimal minimum SINR $\tau^*_{\text{relay}}$ is then given by (\ref{tau}) under these expressions of $\textbf{h}_k$ and $\textbf{V}$. For the HD-AF relay, the minimum user rate is given as $\frac{1}{2} \log_2 (1+\tau^*_{\text{relay}})$, where as for FD-AF relay, we assume a negligible SI value and therefore the minimum user rate is given as $\log_2 (1+\tau^*_{\text{relay}})$.

 We note from Fig. \ref{multiuser_fullrank_opt} that RIS-assisted system with only $N=15$ passive reflecting elements can outperform HD-AF relay where as with $N=180$ can outperform FD-AF relay. When the reflection coefficient is reduced to $\alpha=.8$, $N=22$ and $270$ elements are needed to outperform HD-AF and FD-AF relays respectively. Note that the proposed algorithm is sub-optimal and therefore even higher gains can be realized, with more efficient reflect beamforming designs in the future. Provided low-cost and low-power large RISs are made available in the future, which is expected given the advancements being made in the design of reconfigurable meta-surfaces \cite{Renzo, relay_p1}, the BS will be able to serve blocked users by relying on a passive reflecting surface instead of an active relay, resulting in cheaper implementation and lower energy consumption.



 \vspace{-.15in}
\section{Conclusion}

RIS-assisted MIMO communication is envisioned to achieve unprecedented spectral efficiency gains using only passive reflecting elements at the RIS, that induce phase shifts to reconfigure the signal propagation between the BS and the users. In this work, we studied the performance of the OLP that maximizes the minimum SINR subject to a power constraint for any given RIS phase shift matrix. The fixed-point equation that solves for the minimum SINR under the OLP is expressed in a closed-form for the scenario where the BS-to-RIS LoS channel is of rank-one.  However, this scenario is shown to be limited in efficiently serving more than one user. Next we assume that the LoS channel has high rank and resort to tools from RMT to develop  deterministic approximations for the  parameters of OLP. Finally, we design the RIS phase matrix that maximizes the minimum SINR using projected gradient ascent. Numerical results compare the performance of RIS-assisted system with that of AF relays and discuss its potential. 


There are several challenges that exist in the design and analysis of RIS-assisted systems. The development of channel estimation protocols while maintaining the passive nature of RISs is an immediate challenge that needs to be addressed.  An interesting research direction worth investigating would be to study the asymptotic performance of RIS-assisted systems  under a practical imperfect CSI model and to analyze the susceptibility of these systems to channel estimation errors. The impact on the network latency of estimating all the RIS-assisted channels as well synchronizing the RIS operation with the BS and the users, needs to be studied. Another important research direction would be to study  the more general case with multiple RISs, which can result in higher-rank channels. The joint transmit and reflect beamforming design subject to practical discrete phase-level constraints at the RIS is another direction in which this work can be extended. Moreover, judiciously deploying RISs in a wireless network, comprising both active BSs and passive RISs,  to optimize its performance is another crucial problem to solve. 

 \vspace{-.1in}

\appendices
\section{Proof of Theorem 1}
The proof starts by writing,
\begin{align}
\label{tau4}
&\tau^*=\frac{KP_{max}}{\sum_{k=1}^{K}\frac{1}{d_{k}^*}}, \text{and }q_{k}^*=\frac{\tau^*}{d_{k}^*}, \\
\label{d_defff}
&\text{where, } d_{k}^*=\frac{1}{K}\textbf{h}_{k}^H\left(\sum_{i\neq k} \frac{q_{i}^*}{K}\textbf{h}_{i}\textbf{h}_{i}^{H}+\textbf{I}_{M}\right)^{-1}\textbf{h}_{k}=\frac{1}{K}\textbf{h}_{k}^H\left(\sum_{i\neq k} \frac{\tau^*}{Kd_{i}^*}\textbf{h}_{i}\textbf{h}_{i}^{H}+\textbf{I}_{M}\right)^{-1}\textbf{h}_{k}. 
\end{align}
Using the definition of $\textbf{h}_{k}$, we have,
\begin{align}
\label{wood1}
&d_{k}^*=\frac{1}{K}\rho_{k}\textbf{h}_{2,k}^H \textbf{R}_{RIS_k}^{1/2} \boldsymbol{\Phi}^{H} \textbf{b}\textbf{a}^{H} \left(\alpha_{k}\textbf{a}\textbf{a}^{H}+\textbf{I}_{M}\right)^{-1}\textbf{a}\textbf{b}^{H}\boldsymbol{\Phi} \textbf{R}_{RIS_k}^{1/2} \textbf{h}_{2,k}, \\
\label{alpha2}
&\text{where, } \alpha_{k}=\sum_{i\neq k} \frac{\tau^*}{Kd_{i}^*}\rho_i \textbf{b}^{H}\boldsymbol{\Phi} \textbf{R}_{RIS_i}^{1/2} \textbf{h}_{2,i}\textbf{h}_{2,i}^{H} \textbf{R}_{RIS_i}^{1/2} \boldsymbol{\Phi}^{H} \textbf{b}. 
\end{align}
Using Woodbury formula and that $\textbf{a}^{H}\textbf{a}=M$ (see definition of \textbf{a} in Section V) we have,
\begin{align}
\label{wood}
&\left(\alpha_{k}\textbf{a}\textbf{a}^{H}+\textbf{I}_{M}\right)^{-1}=\textbf{I}_{M}-\frac{\alpha_{k}\textbf{a}\textbf{a}^{H}}{1+\alpha_{k}\textbf{a}^{H}\textbf{a}}. \end{align}
\begin{align}
\label{dk}
&d_{k}^*=\frac{M}{K}\rho_k \frac{\textbf{h}_{2,k}^{H} \textbf{R}_{RIS_k}^{1/2} \boldsymbol{\Phi}^{H} \textbf{b} \textbf{b}^{H}\boldsymbol{\Phi} \textbf{R}_{RIS_k}^{1/2} \textbf{h}_{2,k}}{1+M\alpha_{k}}.
\end{align}
Taking the ratio of $\tau^*$ and $d_{k}^*$ and re-arranging both sides, we obtain $\frac{\tau^* \rho_k\textbf{h}_{2,k}^{H} \textbf{R}_{RIS_k}^{1/2} \boldsymbol{\Phi}^{H} \textbf{b} \textbf{b}^{H}\boldsymbol{\Phi} \textbf{R}_{RIS_k}^{1/2} \textbf{h}_{2,k}}{d_{k}^* K}= \frac{\tau^* (1+M\alpha_{k})}{M}$. Taking $\sum_{k\neq l}$ on both sides, we obtain $\alpha_{l}=\sum_{k\neq l} \frac{\tau^* (1+M\alpha_{k})}{M}=\frac{(K-1)\tau^*}{M}+\tau^*\sum_{k\neq l} \alpha_{k}$. Let $\boldsymbol{\alpha}=[\alpha_{1},\dots, \alpha_{K}]^{T}$, then $\boldsymbol{\alpha}=\frac{(K-1)\tau^*}{M}  \textbf{1}_{K}+\tau^*(\textbf{1}_{K}\textbf{1}_{K}^{T}-\textbf{I}_{K}) \boldsymbol{\alpha}$. This can be reduced to $\left[ \textbf{I}_{K} -\frac{\tau^* \textbf{1}_{K}\textbf{1}_{K}^{T}}{1+\tau^*} \right]\boldsymbol{\alpha}=\frac{\tau^* (K-1)}{M(1+\tau^*)} \textbf{1}_{K}$. Applying Woodbury identity on $\left[ \textbf{I}_{K} -\frac{\tau^* \textbf{1}_{K}\textbf{1}_{K}^{T}}{1+\tau^*} \right]^{-1}\\=\textbf{I}_{K} +\frac{\frac{\tau^*}{1+\tau^*} \textbf{1}_{K}\textbf{1}_{K}^{T}}{1-\frac{\tau^*}{1+\tau^*} \textbf{1}_K^{T}\textbf{1}_K}$, will reveal that all $\alpha_{k}$s are identical and are given by,
\begin{align}
\label{alpha}
&\alpha_{k}=\frac{\tau^* (K-1)}{M(1-\tau^*(K-1))}.
\end{align}
Substituting this $\alpha_{k}$ in the expression of $d^*_{k}$ in (\ref{dk}), and using the resulting expression in (\ref{tau4}), would yield the expression of  $\tau^*$ as,
\begin{align}
\label{tau1}
&\tau^*=\frac{KP_{max}}{\sum_{k=1}^{K} \frac{K}{M\rho_k \textbf{h}_{2,k}^{H} \textbf{R}_{RIS_k}^{1/2} \boldsymbol{\Phi}^{H} \textbf{b} \textbf{b}^{H}\boldsymbol{\Phi} \textbf{R}_{RIS_k}^{1/2} \textbf{h}_{2,k}}   \frac{1}{(1-\tau^*(K-1))}}.
\end{align}
Denote $\textbf{Z}= \frac{1}{M}\sum_{k=1}^{K} \frac{1}{\rho_k \textbf{h}_{2,k}^{H} \textbf{R}_{RIS_k}^{1/2} \boldsymbol{\Phi}^{H} \textbf{b} \textbf{b}^{H}\boldsymbol{\Phi} \textbf{R}_{RIS_k}^{1/2} \textbf{h}_{2,k}} $, the expression in (\ref{tau1}) can be simplified to $\frac{\tau^* Z}{1-\tau^*(K-1)}=P_{max}$. Solving it for $\tau^*$ will yield the expression in \textbf{Theorem 1}. Similarly the OLP in (\ref{G_opt}) can be expressed in terms of $d_{i}^*$ as $\textbf{g}_{k}^*=\frac{\left(\sum_{i\neq k} \frac{\tau^*}{K d_{i}^*}\textbf{h}_{i}\textbf{h}_{i}^{H}+\textbf{I}_{M}\right)^{-1}\textbf{h}_{k}}{||\left(\sum_{i\neq k} \frac{\tau^*}{K d_{i}^*}\textbf{h}_{i}\textbf{h}_{i}^{H}+\textbf{I}_{M}\right)^{-1}\textbf{h}_{k}||}$. Expressing $\left(\sum_{i\neq k} \frac{\tau^*}{K d_{i}^*}\textbf{h}_{i}\textbf{h}_{i}^{H}+\textbf{I}_{M}\right)^{-1}$ in terms of $\alpha_{k}$ as done in (\ref{wood1}), where $\alpha_{k}$ is given by (\ref{alpha2}) and applying  Woodbury formula from (\ref{wood}) on it would yield,
\begin{align}
&\left(\sum_{i\neq k} \frac{\tau^*}{K d_{i}^*}\textbf{h}_{i}\textbf{h}_{i}^{H}+\textbf{I}_{M}\right)^{-1}\textbf{h}_{k}=\sqrt{\rho_{k}} \left( \textbf{I}_{M}-\frac{\alpha_{k}\textbf{a}\textbf{a}^{H}}{1+\alpha_{k}M} \right)\textbf{a}\textbf{b}^{H}\boldsymbol{\Phi} \textbf{R}_{RIS_k}^{1/2} \textbf{h}_{2,k}, \\
&=\sqrt{\rho_{k}} \left( \frac{(1+M\alpha_{k})\textbf{a}-M\alpha_{k}\textbf{a}}{1+\alpha_{k}M} \right)\textbf{b}^{H}\boldsymbol{\Phi} \textbf{R}_{RIS_k}^{1/2} \textbf{h}_{2,k}=\frac{1}{(1+M\alpha_{k})}\textbf{h}_{k}.
\end{align}
The OLP is simplified as  $\textbf{g}_{k}^*=\frac{\frac{1}{(1+M\alpha_{k})}\textbf{h}_{k}}{||\frac{1}{(1+M\alpha_{k})}\textbf{h}_{k}||}=\frac{\textbf{h}_{k}}{||\textbf{h}_{k}||}$, completing the proof of \textbf{Theorem 1}.

\textit{Proof of \textbf{Corollary 2}:}
The proof is similar to that of the single RIS setting, except that we will rely on bounds here as exact closed-form expressions can not be derived.  Using the definition of $\textbf{h}_{k}$ from (\ref{ch_ov1}), defining $d_k^*$ as done in (\ref{d_defff}) and applying the Woodbury formula (under the assumption that the LoS channel vectors are orthogonal for different RISs), we obtain,
\begin{align}
&d_{k}^*=\frac{1}{K} \sum_{l=1}^L \frac{M}{1+M\alpha_{l,k}} \rho_{l,k}\textbf{h}_{2,l,k}^{H}  \boldsymbol{\Phi}_l^{H} \textbf{b}_l \textbf{b}_l^{H}\boldsymbol{\Phi}_l  \textbf{h}_{2,l,k}, \\
\label{alpha33}
&\text{where, } \alpha_{l,k}=\sum_{i\neq k} \frac{\tau^*}{Kd_{i}^*} \rho_{l,i} \textbf{b}_l^{H}\boldsymbol{\Phi}_l  \textbf{h}_{2,l,i}\textbf{h}_{2,l,i}^{H} \boldsymbol{\Phi}_l^{H} \textbf{b}_l,
\end{align}
We can upper bound $d_k^*$ as $d_k^*\leq \frac{1}{K}\textbf{h}_k^H \textbf{h}_k$ and lower bound $\alpha_{l,k}$ as $\alpha_{l,k}\geq \frac{1}{K} \sum_{i\neq k} \frac{\tau^*}{\frac{1}{K}\textbf{h}_i^H \textbf{h}_i} \rho_{l,i} \textbf{b}_l^{H}\boldsymbol{\Phi}_l \\ \textbf{h}_{2,l,i}\textbf{h}_{2,l,i}^{H} \boldsymbol{\Phi}_l^{H} \textbf{b}_l$. Using this bound, we can upper bound $d_k^*$ as,
\begin{align}
&d_{k}^*\leq \frac{1}{K} \sum_{l=1}^L\frac{1}{\alpha_{l,k}} \rho_{l,k}\textbf{h}_{2,l,k}^{H}  \boldsymbol{\Phi}_l^{H} \textbf{b}_l \textbf{b}_l^{H}\boldsymbol{\Phi}_l  \textbf{h}_{2,l,k} \leq \frac{1}{K} \sum_{l=1}^L\frac{\rho_{l,k} \textbf{h}_{2,l,k}^{H}  \boldsymbol{\Phi}_l^{H} \textbf{b}_l \textbf{b}_l^{H}\boldsymbol{\Phi}_l  \textbf{h}_{2,l,k} \text{max}_i \frac{1}{K}\textbf{h}_i^H \textbf{h}_i}{\tau^* \frac{1}{K}\textbf{b}_l^{H}\boldsymbol{\Phi}_l \textbf{H}_{2,l[k]} \bar{\textbf{P}}_{l[k]} \textbf{H}_{2,l[k]}^{H} \boldsymbol{\Phi}_l^{H} \textbf{b}_l},\\
&d_k^* \leq \frac{1}{K} L \text{ max}_{l,k}\frac{\rho_{l,k} \textbf{h}_{2,l,k}^{H}  \boldsymbol{\Phi}_l^{H} \textbf{b}_l \textbf{b}_l^{H}\boldsymbol{\Phi}_l  \textbf{h}_{2,l,k} \text{max}_i \frac{1}{K}\textbf{h}_i^H \textbf{h}_i}{\tau^* \frac{1}{K}\textbf{b}_l^{H}\boldsymbol{\Phi}_l \textbf{H}_{2,l[k]}\bar{\textbf{P}}_{l[k]} \textbf{H}_{2,l[k]}^{H} \boldsymbol{\Phi}_l^{H} \textbf{b}_l},
\end{align}
where $\textbf{H}_{2,l[k]}=[\textbf{h}_{2,l,1}, \dots, \textbf{h}_{2,l,k-1}, \textbf{h}_{2,l,k+1}, \dots, \textbf{h}_{2,l,K}]$ and $\bar{\textbf{P}}_{l[k]}=\text{diag}(\rho_{l,1}, \dots, \rho_{l,k-1}, \rho_{l,k+1}, \dots,\\ \rho_{l,K})$. Next we bound $\tau^*$ using its definition in (\ref{tau4}) as $\tau^* \leq P_{max} d_{max}$, where $d_{max}$ is the RHS of above bound on $d_k^*$. This would yield,
\begin{align}
&\tau^* \leq P_{max} \frac{1}{K} L \text{ max}_{l,k}\frac{\rho_{l,k}\textbf{h}_{2,l,k}^{H}  \boldsymbol{\Phi}_l^{H} \textbf{b}_l \textbf{b}_l^{H}\boldsymbol{\Phi}_l  \textbf{h}_{2,l,k} \text{max}_i \frac{1}{K}\textbf{h}_i^H \textbf{h}_i}{\tau^* \frac{1}{K}\textbf{b}_l^{H}\boldsymbol{\Phi}_l \textbf{H}_{2,l[k]} \bar{\textbf{P}}_{l[k]} \textbf{H}_{2,l[k]}^{H} \boldsymbol{\Phi}_l^{H} \textbf{b}_l}.
\end{align}
Making $\tau^*$ the subject would yield the result in \textbf{Corollary 2}.

\section{Proof of Theorem 2}
Recall the definition of $K\tau^{*u}$ from (\ref{eq1}) given by,
\begin{align}
\label{eq3}
&K\tau^{*u}=\frac{\frac{K P_{max}}{d}}{\frac{P_{max}(K-1)}{d}+ \frac{1}{M}\sum_{k=1}^{K} \tilde{Y}_{k} -c\log K +c\log K}=\frac{\frac{P_{max}}{d}}{\frac{P_{max}}{d}-\frac{P_{max}}{Kd}+\frac{c\log K}{K}+ \frac{\frac{1}{M}\sum_{k=1}^{K} \tilde{Y}_{k} -c\log K}{K} },
\end{align}
where $d=\frac{2}{\rho_{max} \sigma^{2}_{X,max}}$. To study the convergence of $K\tau^{*u}$ we first need to study the asymptotic behavior of the RV $\frac{\frac{1}{M}\sum_{k=1}^{K} \tilde{Y}_{k} -c\log K}{K}$. This can be done by utilizing the result from \textbf{Lemma 1} that $\frac{1}{M}\sum_{k=1}^{K} \tilde{Y}_{k} - c\log K \xrightarrow[M,K\rightarrow \infty]{d} G$, where $G\sim \mathcal{S}_{1}(1,0,0)$, i.e.
\begin{align}
&\underset{M,K\rightarrow \infty}{\text{lim}} \mathbb{P}\left[\frac{1}{M}\sum_{k=1}^{K} \tilde{Y}_{k} - c\log K > t\right] = \mathbb{P}[G>t].
\end{align}
To study $\mathbb{P}\left[\frac{\frac{1}{M}\sum_{k=1}^{K} \tilde{Y}_{k} -c\log K}{K}>t\right]=\mathbb{P}\left[\frac{1}{M}\sum_{k=1}^{K} \tilde{Y}_{k} -c\log K>tK\right]$, we use Lemma 2.11 from \cite{vaart_1998} that says if $X_{n}\xrightarrow[n \rightarrow \infty]{d} X$ then $\text{sup}_{x}|P(X_{n}> x)-P(X> x)|\rightarrow 0$. Therefore we have,
\begin{align}
&\left|\text{lim}_{M,K\rightarrow \infty} \mathbb{P}\left[\frac{1}{M}\sum_{k=1}^{K} \tilde{Y}_{k} -c\log K>tK\right]-\mathbb{P}[G>tK]\right| \rightarrow 0.
\end{align}
Using the property of the CDF of a proper distribution we have $\text{lim}_{K\rightarrow \infty} \mathbb{P}[G>tK] =0$. Therefore $\text{lim}_{M,K\rightarrow \infty}\mathbb{P}\left[\frac{1}{M}\sum_{k=1}^{K} \tilde{Y}_{k} -c\log K>tK\right] \rightarrow 0$. This yields,
\begin{align}
&\frac{\frac{1}{M}\sum_{k=1}^{K} \tilde{Y}_{k} -c\log K}{K} \xrightarrow[M,K\rightarrow \infty]{d} 0.
\end{align}
Now we use this result to study the behavior of $K\tau^{*u}-1$. Using the expression of $K\tau^{*u}$ from (\ref{eq3}), we get $K\tau^{*u}-1=\frac{\frac{P_{max}}{d}}{\frac{P_{max}}{d}-c_{K} }-1=\frac{c_{K}}{\frac{P_{max}}{d}-c_{K}}$, where $c_{K}=\frac{P_{max}}{Kd}-\frac{c\log K}{K}- \frac{\frac{1}{M}\sum_{k=1}^{K} \tilde{Y}_{k} -c\log K}{K}$. We have already proved that $\frac{\frac{1}{M}\sum_{k=1}^{K} \tilde{Y}_{k} -c\log K}{K} \xrightarrow[M,K\rightarrow \infty]{d} 0$. Convergence in distribution to a constant  implies convergence in probability, so $\frac{\frac{1}{M}\sum_{k=1}^{K} \tilde{Y}_{k} -c\log K}{K} \xrightarrow[M,K\rightarrow \infty]{p} 0$. Moreover, $\frac{P_{max}}{Kd}-\frac{c\log K}{K} \xrightarrow[K\rightarrow \infty]{} 0$. Therefore $c_{K}\xrightarrow[M,K\rightarrow \infty]{p} 0$. 

Now provided $\text{lim sup } d<\infty$ (which is ensured by \textbf{Assumption 1} and \textbf{Assumption 2}), we have $K\tau^{*u}-1 \xrightarrow[M,K\rightarrow \infty]{p} 0$. This completes the proof of \textbf{Theorem 2}.


\section{Proof of Theorem 3}
We aim at finding the deterministic equivalents of $q^*_{k}$s, and thereby $\tau^*$. We start by defining $\textbf{Q}_{k}^*=\left( \sum_{i\neq k} \frac{q_{i}^*}{K}\textbf{h}_{i}\textbf{h}_{i}^{H}+\textbf{I}_{M} \right)^{-1}$ and $\tilde{\textbf{h}}_{k}=\rho_{k}^{-1/2} \textbf{h}_{k}$. Then $q^*_{k}$ in (\ref{q}) writes as,
\begin{align}
\label{q_def}
&q_{k}^*=\frac{\tau^*}{\frac{\rho_{k}}{K}\tilde{\textbf{h}}_{k}^{H}\textbf{Q}_{k}^*\tilde{\textbf{h}}_{k}},
\end{align}
 Intuitively, from rank-one perturbation lemma (See \textbf{Lemma 3}), all $d_{k}=\frac{1}{K}\tilde{\textbf{h}}_{k}^{H}\textbf{Q}_{k}^*\tilde{\textbf{h}}_{k}$ present the same asymptotic behavior and should converge to the same limit. In light of this observation, we will  focus on the study of the convergence of $d_{k}$s. The convergence of $q_{k}^*$s to $\bar{q}_{k}$s will then follow. Using the definition of $\textbf{Q}_{k}^*$, we can write $d_{k}$ as,
\begin{align}
&d_{k}=\frac{1}{K}\tilde{\textbf{h}}_{k}^{H}\textbf{Q}_{k}^*\tilde{\textbf{h}}_{k}=\frac{1}{K}\tilde{\textbf{h}}_{k}^{H}\left( \sum_{i\neq k} \frac{\tau^*}{Kd_{i}}\tilde{\textbf{h}}_{i}\tilde{\textbf{h}}_{i}^{H}+ \textbf{I}_{M} \right)^{-1} \tilde{\textbf{h}}_{k}.
\end{align}
 Note that the direct application of standard RMT tools to the quadratic form arising in  the expressions of $d_{k}$s is not analytically correct since the coefficients $d_{i}$s and $\tau^*$ are both functions of the channel vectors $\tilde{\textbf{h}}_{k}$s. However, one would expect the coefficients $d_{i}$, $i\neq k$ to be only weakly  dependent on $\tilde{\textbf{h}}_{k}$, and thus considering them as deterministic, although not properly correct, would lead to infer about their asymptotic behavior \cite{MMS, COUILLET201556}. Based on these intuitive arguments and using the results of \textbf{Lemma 4} and \textbf{Lemma 5}, when all $d_k$s are replaced by the same quantity $\tilde{d}$, one could claim that $d_{k}$ must satisfy the following convergence,
\begin{align}
\label{cond}
&\text{max}_{k}|d_{k}/\tilde{d}-1| \rightarrow 0,
\end{align}
where $\tilde{d}$ is given as the unique solution to,
\begin{align}
\label{dtilde}
&\tilde{d}=\frac{1}{K}\text{tr }\textbf{R}\left(\frac{\tau^*}{\tilde{d}(1+\tau^*)}\textbf{R}+ \textbf{I}_{M} \right)^{-1},
\end{align}
where $\textbf{R}=\textbf{H}_{1}\boldsymbol{\Phi} \textbf{R}_{RIS}\boldsymbol{\Phi}^H \textbf{H}_1^H $. The guess is obtained by using the `almost' independence of $d_{i}$, $i\neq k$ from $\tilde{\textbf{h}}_{k}$ to apply \textbf{Lemma 4} on the quadratic term in $d_{k}$. Next let $b_{i}=\frac{\tau^*}{d_{i}}$ and apply \textbf{Lemma 5} to obtain the guess in (\ref{dtilde}). The uniqueness of the solution of the fixed point equation in (\ref{dtilde}) is easily verifiable. We will now provide a rigorous proof for (\ref{cond}). 

Define $e_{k}=\frac{d_{k}}{\tilde{d}}$ and assume $e_{1} < \dots < e_{K}$. Also note that $\textbf{h}_k=\textbf{R}^{1/2} \textbf{z}_k$, where $\textbf{z}_k\sim \mathcal{CN}(\textbf{0},\textbf{I}_M)$. Then $d_{k} =\frac{1}{K}\textbf{z}_{k}^{H} \textbf{R}^{{1/2}^{H}}\left( \sum_{i\neq k} \frac{\tau^*}{K e_{i} \tilde{d}} \textbf{R}^{1/2} \textbf{z}_{i} \textbf{z}_{i}^{H} \textbf{R}^{{1/2}^{H}}+ \textbf{I}_{M} \right)^{-1} \textbf{R}^{1/2} \textbf{z}_{k}$. Divide both sides by $\tilde{d}$ and write the equation for $k=K$ to get,
\begin{align}
&e_{K} \leq \frac{1}{K}\textbf{z}_{K}^{H} \textbf{R}^{{1/2}^{H}}\left( \sum_{i\neq K} \frac{\tau^*}{K e_{K}} \textbf{R}^{1/2} \textbf{z}_{i} \textbf{z}_{i}^{H} \textbf{R}^{{1/2}^{H}}+\tilde{d} \textbf{I}_{M} \right)^{-1}  \textbf{R}^{1/2} \textbf{z}_{K}.
\end{align}
To prove $\text{lim sup } e_{K} \leq 1$, we use contradiction. Let $l>0$, such that \text{lim sup }$e_{K}>1+l$. Then
\begin{align}
\label{eqq}
&1 \leq \frac{1}{K}\textbf{z}_{K}^{H} \textbf{R}^{{1/2}^{H}}\left( \sum_{i\neq K} \frac{\tau^*}{K} \textbf{R}^{1/2} \textbf{z}_{i}\textbf{z}_{i}^{H} \textbf{R}^{{1/2}^{H}}+\tilde{d}(1+l) \textbf{I}_{M} \right)^{-1}  \textbf{R}^{1/2} \textbf{z}_{K}.
\end{align}
We can easily check that $\frac{\tilde{d}(1+l) }{\tau^*}$ stays almost surely in a bounded interval, by noting from  (\ref{dtilde}) that $\tilde{d}\leq \frac{MR}{K}$, where $R=\underset{N}{\text{lim sup}} ||\textbf{R}||$. Therefore we can apply \textbf{Lemma 4} and \textbf{5} to obtain $\frac{1}{K}\textbf{z}_{K}^{H} \textbf{R}^{{1/2}^{H}}\left( \sum_{i\neq K} \frac{\tau^*}{K}\textbf{R}^{1/2} \textbf{z}_{i}\textbf{z}_{i}^{H} \textbf{R}^{{1/2}^{H}}+\tilde{d}(1+l)  \textbf{I}_{M} \right)^{-1} \textbf{R}^{1/2} \textbf{z}_{K} - \mu \xrightarrow[n \rightarrow \infty]{a.s.} 0$, where $\mu=\frac{1}{K}\text{tr }\textbf{R}\left( \frac{\tau^* }{(1+\mu \tau^*)}\textbf{R}+\tilde{d}(1+l)\textbf{I}_{M} \right)^{-1}$, which can be written equivalently as,
\begin{align}
\label{mu1}
&1=\frac{1}{K}\text{tr }\textbf{R}\left( \mu \frac{\tau^* }{(1+\mu \tau^*)}\textbf{R}+\mu \tilde{d}(1+l) \textbf{I}_{M} \right)^{-1},
\end{align}
Note that the RHS is a decreasing function of $\mu$. From the definition of $\tilde{d}$ in (\ref{dtilde}) we write,
\begin{align}
\label{d11}
&1=\frac{1}{K}\text{tr }\textbf{R}\left( \frac{\tau^* }{(1+\tau^*)}\textbf{R}+ \tilde{d} \textbf{I}_{M} \right)^{-1}.
\end{align}
Now given both (\ref{mu1}) and (\ref{d11}) are equal to one and since $(1+l)>1$ and the RHS of (\ref{mu1}) is a decreasing function in $\mu$, so we must have $\mu<1$. But, a contradiction arises when the system size increases because of the fact that $\mu\geq 1$ from (\ref{eqq}). This proves that $\text{lim sup } e_{K} \leq 1$. The same method can be used to prove that $\text{lim inf } e_{1} \geq 1$. Both these results proves (\ref{cond}).


Note that $\tilde{d}$ is still random due to dependence on $\tau^*$.   Using (\ref{cond}) in the definition of $\tau^*$, we have $\tau^*=\frac{P_{max}}{\frac{1}{K}\sum_{k=1}^{K}\frac{1}{\tilde{d}\rho_{k}}} + o(1)$. Using (\ref{dtilde}) and replacing $\tilde{d}$ by $\frac{\tau^*\frac{1}{K}\sum_{k=1}^K \frac{1}{\rho_{k}}}{P_{max}}$, we finally obtain $\tau^*=\frac{1}{K} \text{tr }\textbf{R}\left(\frac{1}{1+\tau^*}\textbf{R}+\xi \textbf{I}_{M} \right)^{-1} +o(1)$, where $\xi=\frac{1}{P_{max}} \frac{1}{K}\sum_{k=1}^{K}\frac{1}{\rho_{k}}$. Using the above equation, we are tempted to discard the vanishing terms and state that a deterministic equivalent of $\tau^*$ is given by $\bar{\tau}$, the unique solution to,
\begin{align}
&\bar{\tau}=\frac{1}{K} \text{tr }\textbf{R}\left(\frac{1}{1+\bar{\tau}}\textbf{R}+\xi \textbf{I}_{M} \right)^{-1}.
\end{align}
This is indeed true since some tedious calculations involving the resolvent identity: $\textbf{A}^{-1} - \textbf{B}^{-1} = \textbf{A}^{-1} (\textbf{B} - \textbf{A})\textbf{B}^{-1}$, and the inequality: $\frac{1}{K}\text{tr }\textbf{A}\textbf{B}\leq ||\textbf{A}||\frac{1}{K}\text{tr }\textbf{B}$ yield $\tau^*-\bar{\tau}\leq (\tau^*-\bar{\tau})\frac{\bar{\tau}}{1+\bar{\tau}}+o(1)$, from which it follows that $|\tau^*-\bar{\tau}|\rightarrow 0$. Putting the convergence results of $\tau^*$ and $d_{k}$s together in (\ref{q_def}), the convergence of $q^*_{k}$s follows. This completes the proof of \textbf{Theorem 3}.

\section{Proof of Lemma 2}
The computation of the gradient of $\bar{\tau}$ with respect to $\phi_{n}$, $n=1,\dots, N$ will require the use of the implicit function theorem on $g$ defined as $g(\bar{\tau}, \boldsymbol{\Phi})=\bar{\tau}-\frac{1}{K} \text{tr }\textbf{H}_{1}\boldsymbol{\Phi}\textbf{R}_{RIS}\boldsymbol{\Phi}^{H}\textbf{H}_{1}^H \textbf{T}$, where $\textbf{T}=\left(\frac{1}{1+\bar{\tau}}\textbf{H}_{1}\boldsymbol{\Phi}\textbf{R}_{RIS}\boldsymbol{\Phi}^{H}\textbf{H}_{1}^H+\xi \textbf{I}_{M} \right)^{-1}$. The implicit function theorem says,
\begin{align}
\label{imp}
&\frac{\partial \bar{\tau}}{\partial \phi_{n}}=-\frac{\partial g}{\partial \phi_{n}}\big{/}\frac{\partial g}{\partial \bar{\tau}}, \\
\label{ft}
&\text{where, } \frac{\partial g}{\partial \bar{\tau}}=1+(1/K) \text{tr }\textbf{H}_{1}\boldsymbol{\Phi}\textbf{R}_{RIS}\boldsymbol{\Phi}^{H}\textbf{H}_{1}^H \textbf{T} \left(-\textbf{H}_{1}\boldsymbol{\Phi}\textbf{R}_{RIS}\boldsymbol{\Phi}^{H}\textbf{H}_{1}^H/(1+\bar{\tau})^2  \right) \textbf{T}.
\end{align}
To compute the expression of $\frac{\partial g}{\partial \phi_{n}}$, we use the fact that,
\begin{align}
\label{step2}
&\frac{\partial g}{\partial \phi_{n}}=2 \frac{\partial g}{\partial \phi_{n}^*}.
\end{align}
Next we compute $\frac{\partial g}{\partial \phi_{n}^*}=-\frac{1}{K} \sum_{i}\sum_{k} \left\{\textbf{H}_{1}\boldsymbol{\Phi}\textbf{R}_{RIS}\boldsymbol{\Phi}^{H}\textbf{H}_{1}^H\right\}_{k,i} \frac{\partial}{\partial \phi_{n}^*} \left\{\textbf{T}\right\}_{i,k}-\frac{1}{K}  \sum_{i}\sum_{k} \left\{\textbf{T}\right\}_{i,k} \\ \times \frac{\partial}{\partial \phi_{n}^*} \{\textbf{H}_{1}\boldsymbol{\Phi}\textbf{R}_{RIS}\boldsymbol{\Phi}^{H}\textbf{H}_{1}^H\}_{k,i}=T_{1}+T_{2}$.  We will deal with $T_{1}$ first as,
\begin{align}
&T_1=\frac{1}{K} \sum_{i}\sum_{k}\sum_{l}\sum_{s} \{\textbf{H}_{1}\boldsymbol{\Phi}\textbf{R}_{RIS}\boldsymbol{\Phi}^{H}\textbf{H}_{1}^H\}_{k,i} \left\{\textbf{T}\right\}_{i,l} \frac{1}{1+\bar{\tau}} \frac{\partial}{\partial \phi_{n}^*} \{\textbf{H}_{1}\boldsymbol{\Phi}\textbf{R}_{RIS}\boldsymbol{\Phi}^{H}\textbf{H}_{1}^H\}_{l,s}  \left\{\textbf{T}\right\}_{s,k}, \nonumber \\
\label{step}
&=\frac{\alpha}{K(1+\bar{\tau})}\Big[\textbf{H}_{1}^H \textbf{T} \textbf{H}_{1}\boldsymbol{\Phi}\textbf{R}_{RIS}\boldsymbol{\Phi}^{H}\textbf{H}_{1}^H  \textbf{T} \textbf{H}_{1}\boldsymbol{\Phi}\textbf{R}_{RIS} \Big]_{n,n}.
\end{align}
The result in (\ref{step}) is obtained by noting  $\frac{\partial}{\partial \phi_{n}^*} \{\textbf{H}_{1}\boldsymbol{\Phi}\textbf{R}_{RIS}\boldsymbol{\Phi}^{H}\textbf{H}_{1}^H\}_{l,s}$=$\alpha [\textbf{H}_{1} \boldsymbol{\Phi} \textbf{R}_{RIS}]_{l,n}[\textbf{H}_{1}^H]^T_{s,n}$. Similar analysis would yield $T_2=-\frac{\alpha}{K}\left[\textbf{H}_1^H \textbf{T} \textbf{H}_{1} \boldsymbol{\Phi} \textbf{R}_{RIS}  \right]_{n,n}$. Putting $T_1$ and $T_2$ together yields,
\begin{align}
&\frac{\partial g}{\partial \phi_{n}^*}=\frac{\alpha}{K}\left[\frac{1}{1+\bar{\tau}} \textbf{H}_1^H\textbf{T} \textbf{H}_1 \boldsymbol{\Phi} \textbf{R}_{RIS} \boldsymbol{\Phi}^H \textbf{H}_1^H \textbf{T} \textbf{H}_1 \boldsymbol{\Phi} \textbf{R}_{RIS} - \textbf{H}_1^H\textbf{T} \textbf{H}_1 \boldsymbol{\Phi} \textbf{R}_{RIS} \right]_{n,n}.
\end{align}
Now using this result in (\ref{step2}) and plugging it and (\ref{ft}) in (\ref{imp}) will yield the expression of $\frac{\partial \bar{\tau}}{\partial \phi_{n}}$.

\section{Useful Lemmas}
Let $\textbf{H}=[\textbf{h}_{1}, \dots , \textbf{h}_{K}]$ be an $M\times K$ matrix, with $\textbf{h}_{k}=\textbf{R}_{k}^{1/2}\textbf{z}_{k}$, where $\textbf{R}_{k}=\textbf{H}_{1}\boldsymbol{\Phi}\textbf{R}_{RIS_{k}}\boldsymbol{\Phi}^H \textbf{H}_1^H$ and $\textbf{z}_k\sim\mathcal{CN}(\textbf{0},\textbf{I}_M)$. Define the resolvent matrix of $\textbf{H}\textbf{B}\textbf{H}^{H}$, where $\textbf{B}=\text{diag}(b_{1},\dots, b_{K})$, as,
\begin{align}
&\textbf{Q}(\sigma^2)=\left(\frac{1}{K}\textbf{H}\textbf{B}\textbf{H}^H+\sigma^2\textbf{I}_{M}  \right)=\left(\frac{1}{K}\sum_{i=1}^{K} b_{i}\textbf{h}_{i}\textbf{h}_{i}^H+\sigma^2\textbf{I}_{M}  \right).
\end{align}
 Define also $\textbf{Q}_{k}(\sigma^2)=\left(\frac{1}{K}\sum_{i\neq k}^{K} b_{i}\textbf{h}_{i}\textbf{h}_{i}^H+\sigma^2\textbf{I}_{M}  \right)$. The following lemmas recall some classical identities involving the resolvent matrix, which have been extensively used in our derivations:

\textbf{Lemma 3.} Rank-one perturbation lemma: For any matrix \textbf{A}, $\text{tr }\textbf{A}(\textbf{Q}(\sigma^2)-\textbf{Q}_{k}(\sigma^2))\leq ||\textbf{A}||_2$.

\textbf{Lemma 4.} (Convergence of quadratic forms)  Let $\textbf{y}\sim \mathcal{CN}(\textbf{0}_{M},\textbf{I}_{M})$. Let \textbf{A} be an $M\times M$ matrix independent of \textbf{y}, which has a bounded spectral norm. Then $\frac{1}{M}\textbf{y}^H\textbf{A}\textbf{y}-\frac{1}{M}\text{tr}(\textbf{A}) \xrightarrow[n\rightarrow\infty]{a.s.} 0$.

\textbf{Lemma 5} (Using \textbf{Theorem 1} from \cite{SINRdeterministic}) Let $m(\sigma^2)$ be the unique solution to,
\begin{align}
&m(\textbf{A}, \sigma^2)=\frac{1}{K}\text{tr}\textbf{A}\left( \frac{1}{K} \sum_{j=1}^{K} \frac{b_{j}\textbf{R}_{j}}{1+e_{j}(\sigma^2)}   +\sigma^2\textbf{I}_{M} \right), 
\end{align}
where $e_{k}(\sigma^2)$ is the unique solution to $e_{k}(\sigma^2)=\frac{1}{K}\text{tr}b_{k}\textbf{R}_{k} \left( \frac{1}{K} \sum_{j=1}^{K} \frac{b_{j}\textbf{R}_{j}}{1+e_{j}(\sigma^2)}   +\sigma^2\textbf{I}_{M} \right)$. 

Consider the asymptotic regime outlined in \textbf{Assumption 3}. Let $[a, b]$ be a closed bounded interval in $[0,\infty]$, then the following convergence
holds true:
\begin{align}
&\text{sup}_{\sigma^2\in[a,b]}\left| \frac{1}{K}\text{tr } \textbf{Q}(\sigma^2)-m(\textbf{I}_{M},\sigma^2) \right| \xrightarrow[n\rightarrow\infty]{a.s.} 0.
\end{align}
Using the rank-one  perturbation lemma $\text{sup}_{\sigma^2\in[a,b]}|\frac{1}{K}\text{tr } \textbf{Q}_{k}(\sigma^2)-m(\textbf{I}_{M},\sigma^2)| \xrightarrow[n\rightarrow\infty]{a.s.} 0$.
\vspace{-.14in}
\bibliographystyle{IEEEtran}
\bibliography{bib}

\end{document}